\def\buildmode{0}

\ifcase\buildmode

  \pdfoutput=1
\documentclass[12pt,a4paper]{article}

\usepackage{ifthen} 
\newboolean{pdflatex}
\setboolean{pdflatex}{true} 

\newboolean{articletitles}
\setboolean{articletitles}{true} 

\newboolean{uprightparticles}
\setboolean{uprightparticles}{false} 

\def\paperauthors{LHCb collaboration} 
\def\paperasciititle{First measurement of the forward rapidity dependence of $W$ boson transverse helicity fractions} 
\def\papertitle{First measurement of the forward rapidity dependence of $W$ boson transverse helicity fractions} 
\def\paperkeywords{{High Energy Physics}, {LHCb}} 
\def\papercopyright{\the\year\ CERN for the benefit of the LHCb collaboration} 
\def\paperlicence{CC BY 4.0 licence}
\def\paperlicenceurl{https://creativecommons.org/licenses/by/4.0/}

\newcommand{\ptw}{\ensuremath{p_\mathrm{T}^{W}}\xspace}

\newcommand{\yW}{\ensuremath{y_{W}}\xspace}
\newcommand{\ptm}{\ensuremath{p^{\mu}_\mathrm{T}}\xspace}
\newcommand{\cost}{\ensuremath{\cos\theta^{*}}\xspace}

\newcommand{\etam}{\ensuremath{\eta^{\mu}}\xspace}
\newcommand*{\MG}{\textsc{MadGraph5}\_aMC@NLO\xspace}

\newif\ifEnableSectionTOCLinks
\EnableSectionTOCLinksfalse 

\usepackage[top=1in, bottom=1.25in, left=1in, right=1in]{geometry}

\usepackage{microtype}
\usepackage{lineno}  
\usepackage{xspace} 
\usepackage{caption} 

\usepackage{graphicx}  
\usepackage{color}
\usepackage{colortbl}
\graphicspath{{./figs/}} 

\usepackage{amsmath} 
\usepackage{amssymb}
\usepackage{amsfonts}
\usepackage{upgreek} 

\usepackage[normalem]{ulem} 

\newcommand*\patchAmsMathEnvironmentForLineno[1]{%
\expandafter\let\csname old#1\expandafter\endcsname\csname #1\endcsname
\expandafter\let\csname oldend#1\expandafter\endcsname\csname
end#1\endcsname
 \renewenvironment{#1}%
   {\linenomath\csname old#1\endcsname}%
   {\csname oldend#1\endcsname\endlinenomath}%
}
\newcommand*\patchBothAmsMathEnvironmentsForLineno[1]{%
  \patchAmsMathEnvironmentForLineno{#1}%
  \patchAmsMathEnvironmentForLineno{#1*}%
}
\AtBeginDocument{%
\patchBothAmsMathEnvironmentsForLineno{equation}%
\patchBothAmsMathEnvironmentsForLineno{align}%
\patchBothAmsMathEnvironmentsForLineno{flalign}%
\patchBothAmsMathEnvironmentsForLineno{alignat}%
\patchBothAmsMathEnvironmentsForLineno{gather}%
\patchBothAmsMathEnvironmentsForLineno{multline}%
\patchBothAmsMathEnvironmentsForLineno{eqnarray}%
}

\usepackage[pdftex,
            pdfauthor={\paperauthors},
            pdftitle={\paperasciititle},
            pdfkeywords={\paperkeywords}]{hyperref}
\usepackage{hyperxmp}
\hypersetup{
    pdfcopyright={Copyright (C) \papercopyright},
    pdflicenseurl={\paperlicenceurl}
}

\usepackage[colorinlistoftodos,textsize=scriptsize]{todonotes}

\usepackage[bottom,flushmargin,hang,multiple]{footmisc}

\usepackage[all]{hypcap} 

\usepackage{xspace} 
\usepackage{upgreek}

\def\lhcb   {\mbox{LHCb}\xspace}

\def\lhc    {\mbox{LHC}\xspace}

\def\tevatron {Tevatron\xspace}

\def\MagUp {\mbox{\em Mag\kern -0.05em Up}\xspace}

\ifdefined\ifuprightparticles
\else
\newboolean{uprightparticles}
\setboolean{uprightparticles}{false} 
\fi

\ifthenelse{\boolean{uprightparticles}}%
{

 \def\Pmu         {\ensuremath{\upmu}\xspace}

 \def\Ppsi        {\ensuremath{\uppsi}\xspace}

 \def\PDelta      {\ensuremath{\Delta}\xspace}                 
 \def\PXi         {\ensuremath{\Xi}\xspace}                 
 \def\PLambda     {\ensuremath{\Lambda}\xspace}                 
 \def\PSigma      {\ensuremath{\Sigma}\xspace}                 
 \def\POmega      {\ensuremath{\Omega}\xspace}                 
 \def\PUpsilon    {\ensuremath{\Upsilon}\xspace}
 \let\oldPi\Pi
 \def\PPi         {\ensuremath{\oldPi}\xspace}

 \def\PB      {\ensuremath{\mathrm{B}}\xspace}                 
 \def\PD      {\ensuremath{\mathrm{D}}\xspace}                 
 \def\PJ      {\ensuremath{\mathrm{J}}\xspace}                 
 \def\PK      {\ensuremath{\mathrm{K}}\xspace}                 
 \def\PW      {\ensuremath{\mathrm{W}}\xspace}                 
 \def\PZ      {\ensuremath{\mathrm{Z}}\xspace}                 
 \def\Ps      {\ensuremath{\mathrm{s}}\xspace}

 \def\thebaroffset{0.0em}
}
{

 \def\Pmu         {\ensuremath{\mu}\xspace}

 \def\Ppsi        {\ensuremath{\psi}\xspace}                 
                  
 \mathchardef\PDelta="7101
 \mathchardef\PXi="7104
 \mathchardef\PLambda="7103
 \mathchardef\PSigma="7106
 \mathchardef\POmega="710A
 \mathchardef\PUpsilon="7107
 \mathchardef\PPi="7105
 \def\PB      {\ensuremath{B}\xspace}                 
 \def\PD      {\ensuremath{D}\xspace}                 
 \def\PJ      {\ensuremath{J}\xspace}                 
 \def\PK      {\ensuremath{K}\xspace}                 
 \def\PW      {\ensuremath{W}\xspace}                 
 \def\PZ      {\ensuremath{Z}\xspace}                 
 \def\Ps      {\ensuremath{s}\xspace}

 \def\thebaroffset{0.18em}
}
\newcommand{\offsetoverline}[2][\thebaroffset]{\kern #1\overline{\kern -#1 #2}}%

\makeatletter
\ifcase \@ptsize \relax
  \newcommand{\miniscule}{\@setfontsize\miniscule{4}{5}}
\or
  \newcommand{\miniscule}{\@setfontsize\miniscule{5}{6}}
\or
  \newcommand{\miniscule}{\@setfontsize\miniscule{5}{6}}
\fi
\makeatother

\DeclareRobustCommand{\optbar}[1]{\shortstack{{\miniscule (\rule[.5ex]{1.25em}{.18mm})}
  \\ [-.7ex] $#1$}}

\def\mumu       {{\ensuremath{\Pmu^+\Pmu^-}}\xspace}

\def\W      {{\ensuremath{\PW}}\xspace}
\def\Wp     {{\ensuremath{\PW^+}}\xspace}
\def\Wm     {{\ensuremath{\PW^-}}\xspace}

\def\Z      {{\ensuremath{\PZ}}\xspace}

\def\squark    {{\ensuremath{\Ps}}\xspace}

\def\KorKbar {\kern \thebaroffset\optbar{\kern -\thebaroffset \PK}{}\xspace}

\def\D       {{\ensuremath{\PD}}\xspace}

\def\DorDbar {\kern \thebaroffset\optbar{\kern -\thebaroffset \PD}\xspace}

\def\Dp      {{\ensuremath{\D^+}}\xspace}
\def\Dm      {{\ensuremath{\D^-}}\xspace}

\def\DpDm    {\ensuremath{\Dp {\kern -0.16em \Dm}}\xspace}

\def\B       {{\ensuremath{\PB}}\xspace}

\def\BorBbar {\kern \thebaroffset\optbar{\kern -\thebaroffset \PB}\xspace}

\def\Bd      {{\ensuremath{\B^0}}\xspace}

\def\BdorBdbar {\kern \thebaroffset\optbar{\kern -\thebaroffset \Bd}\xspace}

\def\Bs      {{\ensuremath{\B^0_\squark}}\xspace}

\def\BsorBsbar {\kern \thebaroffset\optbar{\kern -\thebaroffset \Bs}\xspace}

\def\jpsi     {{\ensuremath{{\PJ\mskip -3mu/\mskip -2mu\Ppsi}}}\xspace}

\def\Y#1S{\ensuremath{\PUpsilon{(#1S)}}\xspace}
\def\OneS  {{\Y1S}\xspace}

\def\LorLbar     {\kern \thebaroffset\optbar{\kern -\thebaroffset \PLambda}\xspace}

\newcommand{\decay}[2]{\ensuremath{\mathinner{#1\!\to #2}}\xspace}

\def\to                 {\ensuremath{\rightarrow}\xspace}

\def\AT#1     {\ensuremath{A_{\mathrm{T}}^{#1}}\xspace}           

\def\C#1      {\ensuremath{\mathcal{C}_{#1}}\xspace}                       
\def\Cp#1     {\ensuremath{\mathcal{C}_{#1}^{'}}\xspace}                    
\def\Ceff#1   {\ensuremath{\mathcal{C}_{#1}^{\mathrm{(eff)}}}\xspace}        
\def\Cpeff#1  {\ensuremath{\mathcal{C}_{#1}^{'\mathrm{(eff)}}}\xspace}       
\def\Ope#1    {\ensuremath{\mathcal{O}_{#1}}\xspace}                       
\def\Opep#1   {\ensuremath{\mathcal{O}_{#1}^{'}}\xspace}                    

\newcommand{\nospaceunit}[1]{\ensuremath{\text{#1}}}       
\newcommand{\aunit}[1]{\ensuremath{\text{\,#1}}}       

\newcommand{\tev}{\aunit{Te\kern -0.1em V}\xspace}
\newcommand{\gev}{\aunit{Ge\kern -0.1em V}\xspace}
\newcommand{\mev}{\aunit{Me\kern -0.1em V}\xspace}
\newcommand{\kev}{\aunit{ke\kern -0.1em V}\xspace}
\newcommand{\ev}{\aunit{e\kern -0.1em V}\xspace}
 
\newcommand{\mevc}{\ensuremath{\aunit{Me\kern -0.1em V\!/}c}\xspace}
\newcommand{\gevc}{\ensuremath{\aunit{Ge\kern -0.1em V\!/}c}\xspace}
\newcommand{\mevcc}{\ensuremath{\aunit{Me\kern -0.1em V\!/}c^2}\xspace}
\newcommand{\gevcc}{\ensuremath{\aunit{Ge\kern -0.1em V\!/}c^2}\xspace}

\def\mum  {\ensuremath{\,\upmu\nospaceunit{m}}\xspace}

\def\fb   {\ensuremath{\aunit{fb}}\xspace}
\def\invfb   {\ensuremath{\fb^{-1}}\xspace}

\def\gsim{{~\raise.15em\hbox{$>$}\kern-.85em
          \lower.35em\hbox{$\sim$}~}\xspace}
\def\lsim{{~\raise.15em\hbox{$<$}\kern-.85em
          \lower.35em\hbox{$\sim$}~}\xspace}

\def\pt         {\ensuremath{p_{\mathrm{T}}}\xspace}

\def\powheg     {\mbox{\textsc{Powheg}}\xspace}

\def\tell1  {TELL1\xspace}
\def\ukl1   {UKL1\xspace}

\newcommand{\ie}{\mbox{\itshape i.e.}\xspace}

\newcommand{\lhcborcid}[1]{\href{https://orcid.org/#1}{\hspace*{0.1em}\raisebox{-0.45ex}{\includegraphics[width=1em]{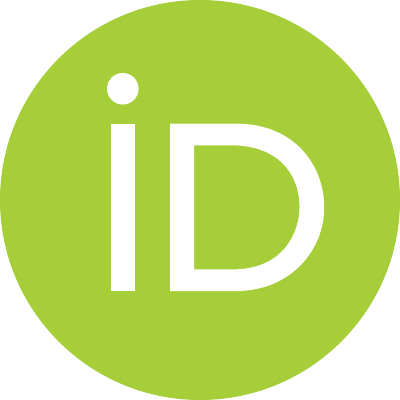}}}}

\hypersetup{
  colorlinks   = true, 
  urlcolor     = blue, 
  linkcolor    = blue, 
  citecolor    = red   
}

\ifEnableSectionTOCLinks
    \usepackage[explicit]{titlesec} 
    
    \let\oldcontentsline\contentsline
    \renewcommand

    \titleformat{\section}{\normalfont\Large\bf}{\hyperlink{tocsection.\thesection}{{\thesection} \parbox[t]{\dimexpr\textwidth-1pc}{#1}}}{1pc}{}

    \titleformat{\subsection}{\normalfont\bf}{\hyperlink{tocsubsection.\thesubsection}{{\thesubsection} \parbox[t]{\dimexpr\textwidth-1pc}{#1}}}{1pc}{}

    \titleformat{name=\section,numberless}[display]{}{}{0pt}{\normalfont\Huge\bfseries #1}
\fi

\usepackage{cite} 
\usepackage{LHCb/mciteplus}
\makeatletter
\g@addto@macro\bfseries{\boldmath}
\makeatother

\usepackage{longtable} 
\usepackage{xspace}
\usepackage{arydshln}

\begin{document}

\renewcommand{\thefootnote}{\fnsymbol{footnote}}
\setcounter{footnote}{1}


\begin{titlepage}
\pagenumbering{roman}

\vspace*{-1.5cm}
\centerline{\large EUROPEAN ORGANIZATION FOR NUCLEAR RESEARCH (CERN)}
\vspace*{1.5cm}
\noindent
\begin{tabular*}{\linewidth}{lc@{\extracolsep{\fill}}r@{\extracolsep{0pt}}}
\ifthenelse{\boolean{pdflatex}}
{\vspace*{-1.5cm}\mbox{\!\!\!\includegraphics[width=.14\textwidth]{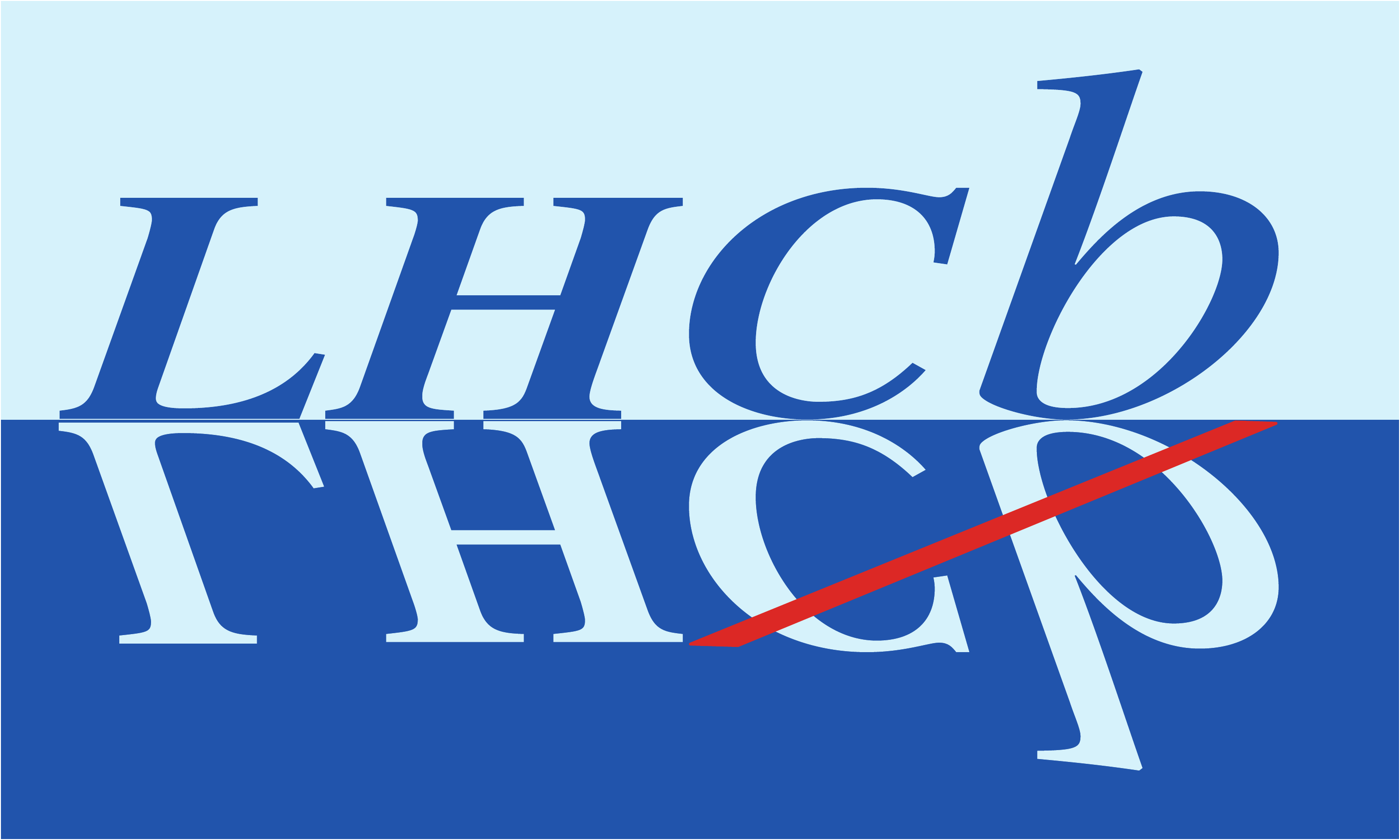}} & &}%
{\vspace*{-1.2cm}\mbox{\!\!\!\includegraphics[width=.12\textwidth]{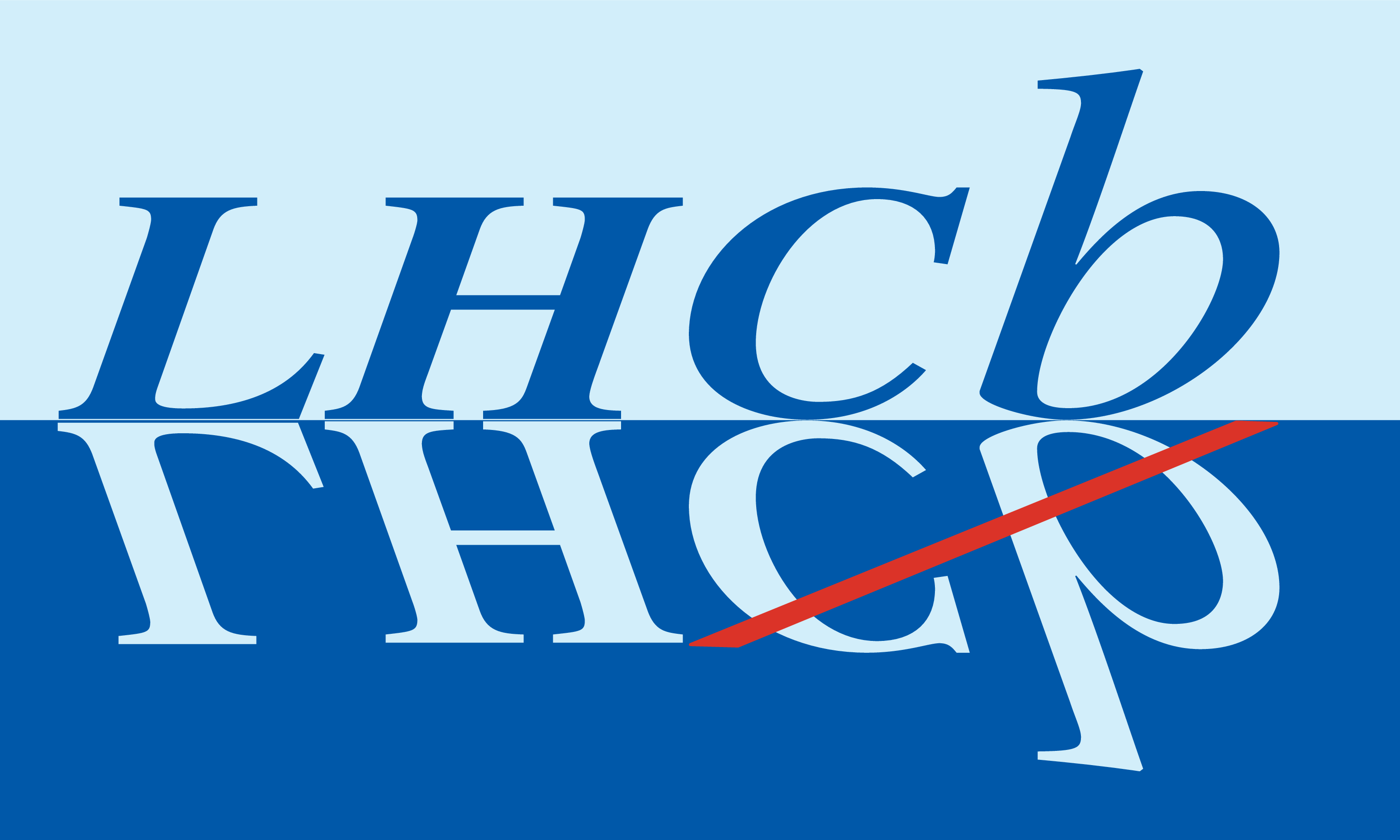}} & &}%
\\
 & & CERN-EP-2026-247 \\  
 & & LHCb-PAPER-2026-035 \\  
 & & 14 November 2026 \\ 
 & & \\
\end{tabular*}

\vspace*{4.0cm}

{\normalfont\bfseries\boldmath\huge
\begin{center}
  \papertitle 
\end{center}
}

\vspace*{2.0cm}

\begin{center}
\paperauthors\footnote{Authors are listed at the end of this Letter.}
\end{center}

\vspace{\fill}

\begin{abstract}
  \noindent
The transverse helicity fractions of $W$ bosons are measured as a function of the $W$ boson rapidity, $\yW$, in the range $0 \leq  \yW \leq 5$ using $W\to\mu\nu_{\mu}$ decays in $pp$ collisions at $\sqrt{s}$ = 13\tev recorded by the \lhcb experiment and corresponding to an integrated luminosity of $5.1\invfb$. The fractions are extracted from a template fit to the muon transverse momentum and pseudorapidity. The results show a strong rapidity dependence and agree with next-to-leading-order Standard Model predictions, providing the first determination of the transverse helicity fractions of \W bosons in the forward region.

\end{abstract}

\vspace*{2.0cm}

\begin{center}
  Submitted to
  Phys.~Rev.~Lett. 

\end{center}

\vspace{\fill}

{\footnotesize 
\centerline{\copyright~\papercopyright. \href{\paperlicenceurl}{\paperlicence}.}}
\vspace*{2mm}

\end{titlepage}


\newpage
\setcounter{page}{2}
\mbox{~}
%
%
%
%


\renewcommand{\thefootnote}{\arabic{footnote}}
\setcounter{footnote}{0}


\cleardoublepage


\pagestyle{plain} 
\setcounter{page}{1}
\pagenumbering{arabic}



In the Standard Model (SM), the charged-current weak interaction has a chiral ($V$--$A$) structure in which \W bosons couple to left-handed fermions and right-handed antifermions~\cite{GLASHOW1961579,PhysRevLett.19.1264,SALAM1964168}. According to the projection of the \W boson spin along its direction of motion, three helicity states are allowed: two transverse states, labeled as right- and left-handed with the spin aligned parallel or antiparallel to the direction of motion respectively, and one longitudinal state with the spin perpendicular to the direction of motion. At hadron colliders, the chiral structure of the weak interaction combines with the initial-state parton momentum distributions~\cite{Bjorken:1969ja} to produce characteristic transverse helicity patterns of \W bosons from tree-level quark-antiquark annihilation. 
Higher-order QCD contributions introduce a longitudinal component and modify the transverse fractions depending on the \W boson transverse momentum, \ptw, where \pt is the component of the momentum transverse to the beam, and rapidity, \yW~\cite{Bern:2011ie}.

The polarization of the \W boson, meaning the relative population of these helicity states, has been studied at the \tevatron ~\cite{CDF:2008eba,D0:2010zpi}, with the PHENIX experiment~\cite{PHENIX:2015ade}, and the \lhc~\cite{ATLAS:2012nhi,CMS:2014uod} through top quark decays, production at high \ptw~\cite{CMS:2011kaj,ATLAS:2012au}, and direct measurements of rapidity-dependent \W helicity distributions~\cite{CMS:2020cph}.  
The forward acceptance of the \lhcb detector, covering the pseudorapidity range $2<\eta<5$, provides access to a previously unexplored region where the partonic collisions are highly asymmetric. For these collisions, a particularly pronounced separation between the left- and right-handed \W helicity fractions is predicted.

This Letter reports the first measurement of the \W boson transverse helicity fractions in the forward region, using $\W\to\mu\nu_{\mu}$ decays recorded with the \lhcb detector in proton--proton, $pp$, collisions at a center-of-mass energy of 13\tev during 2016--2018 and corresponding to an integrated luminosity of $5.1\invfb$. Their rapidity dependence tests the interplay between parton distribution functions, PDFs, higher-order QCD effects, and the left-handed coupling of the weak interaction.

Because the muon transverse momentum, $\ptm$, and pseudorapidity, $\etam$, distributions depend on the \W-boson helicity state, they can be used to separate the contributing helicity states. The left-handed, $f_{\mathrm{L}}$, and right-handed, $f_{\mathrm{R}}$, fractions are measured as a function of \yW for \Wp and \Wm bosons, while the longitudinal fraction, $f_{\mathrm{0}}$, is estimated from the SM prediction as described below.

The \lhcb detector~\cite{LHCb-DP-2008-001,LHCb-DP-2014-002} is a single-arm forward spectrometer covering the pseudorapidity range $2<\eta<5$. The tracking system~\cite{LHCb-DP-2017-001} provides a measurement of the momentum, $p$, of charged particles with a relative uncertainty that varies from 0.5\% at low momentum to 1.0\% at 200$\gev$, where natural units with $c=\hbar$ = 1 are used throughout this Letter, and an impact parameter resolution of $(15+29/\pt)\,\mum$~\cite{LHCb-DP-2014-001}. Muons are identified by alternating layers of iron and multiwire proportional chambers~\cite{LHCb-DP-2012-002}, with misidentification probabilities below the percent level. The online selection uses a hardware trigger requiring a high-\pt muon candidate, then a software trigger requiring a well-reconstructed isolated muon with large~\pt~\cite{LHCb-DP-2012-004}. Triggered data further undergo a centralised, offline processing step to deliver physics-analysis-ready data across the entire LHCb physics program~\cite{Grieser:2025nch}.

Simulation is required to model the effects of the detector acceptance, the trigger and reconstruction efficiencies, the event-selection requirements, and to construct templates for the various \W-boson helicity states and background processes. Simulated signal samples of $\W\to\mu\nu_{\mu}$ decays are produced with \MG~\cite{Alwall:2014hca} interfaced to \textsc{Pythia\,8}~\cite{Sjostrand:2007gs} for parton showering and hadronization.  These samples are used to produce the signal templates without providing event-by-event \W boson helicity information. Additional samples generated with \textsc{Powheg-Box}~\cite{Alioli:2010xd} plus \textsc{Pythia\,8}, and with leading-order \textsc{Pythia\,8} are used to assess modeling uncertainties and perform validation studies. Additional samples are generated in a solid angle of 4$\pi$ without the \lhcb fiducial requirements with no detector response. Simulated samples are used to model muon $\ptm$ and $\etam$ distributions of the main background processes. In these samples, $pp$ collisions for different processes are generated using \textsc{Pythia\,8} with a dedicated \lhcb configuration~\cite{LHCb-PROC-2010-056}. Decays of unstable particles are described by \textsc{EvtGen}~\cite{Lange:2001uf}, with final-state photon radiation simulated using \textsc{Photos}~\cite{davidson2015photos}. The interaction of the generated particles with the detector, and the detector response, are implemented using the \textsc{Geant4}~\cite{Agostinelli:2002hh} toolkit as configured in the \lhcb simulation framework. These backgrounds include $Z\to\mu^{+}\mu^{-}$ decays, $\W\to\tau\nu$ decays with $\tau\to\mu\nu\bar\nu$, leptonic decays of top-quark pairs and single-top production in association with a $W$ boson, diboson processes, Drell--Yan dimuon production at low invariant mass, and inclusive production of heavy-flavor hadrons and high-\pt hadrons that mimic isolated muons, further referred to as hadronic mis-ID. 

Muon candidates are reconstructed as tracks with hits in the tracking detectors, associated hits in the muon system, and satisfying tight muon-identification criteria. Candidate muons are required to have  $27\leq \ptm \leq 45\gev$ and  $2.2 \leq \etam \leq 4.4$. Events are required to contain exactly one such muon candidate. The relative uncertainty on the muon momentum, as estimated from the track fit, is required to be below $6\%$ and the track-fit $\chi^{2}$ per degree of freedom must be less than 1.8. The offline selection is aligned with that used in previous \lhcb electroweak measurements of \W boson production~\cite{LHCb-PAPER-2025-071,LHCb-PAPER-2025-070} and of the \W boson mass~\cite{LHCb-PAPER-2021-024}. Data and simulation are processed with identical reconstruction and selection software, and the simulation is subsequently corrected to better reproduce the distributions observed in data as described in the following.

To suppress muons from heavy-flavor and $\W\to\tau\nu$ decays, the signal muons are required to be consistent with originating from a primary vertex. Hadronic backgrounds are reduced using a track-based isolation requirement on the scalar transverse-momentum sum in a cone around the muon~\cite{LHCb-PAPER-2021-024}. The energy deposited in the hadronic calorimeter by the muon is required to be small, further suppressing hadrons that reach the muon system. To reduce contamination from \decay{\Z}{\mumu} decays, events are vetoed if they contain a second track with $\pt > 25\gev$ within $2.0 < \eta < 4.5$, with no explicit muon-identification requirement applied to the track. The overall selection results in 4.5M \Wp and 3.1M \Wm candidates.

The momentum-scale and resolution calibrations follow the procedures established in recent LHCb precision electroweak measurements~\cite{LHCb-PAPER-2021-024,LHCb-PAPER-2024-028,LHCb-PAPER-2025-008}, although the helicity-fraction measurement is largely insensitive to them. Curvature-bias corrections to the reconstructed charge-over-momentum distribution are derived with the pseudomass method in \mbox{\decay{\Z}{\mumu}} decays~\cite{LHCb-DP-2023-001}, using a clean sample with low $\ptm$ selected on the acoplanarity and pseudorapidity of the two muons~\cite{phistar}. The corrections are applied separately for each magnet polarity and muon charge.  In simulation, the muon momentum is smeared to match the data through a transformation comprising energy-loss, momentum-scale, momentum-dependent and -independent smearing, and curvature-bias terms, whose parameters are obtained from a fit to \jpsi and \OneS decays to muon pairs and validated with the \jpsi, \OneS, and \Z mass peaks. Residual mismodeling of the impact-parameter and track-fit-quality distributions is removed by an analogous Gaussian smearing of the simulation, with parameters determined from an unbinned Anderson--Darling test~\cite{anderson1952asymptotic} in intervals of $\etam$ and azimuthal angle of the muon. A correction to the \ptw spectrum, specific to this measurement, is derived from the ratio of the reconstructed \pt distribution of $\Z$ bosons between data and simulation. This ratio, smoothed with a Savitzky--Golay filter~\cite{Savitzky:1964}, is applied to simulated \W events after a leading-order mass rescaling.

The sensitivity to the \W boson helicity arises from the angular distribution of the muon in the \W boson rest frame. The polar angle $\theta^{*}$ is defined in the Collins--Soper frame~\cite{PhysRevD.16.2219} as the angle between the outgoing muon and the bisector of the incoming parton momenta, evaluated in the \W rest frame. Neglecting the muon mass, the normalized differential cross-section can be written~\cite{Ellis_Stirling_Webber_1996}, at leading order, as
\begin{equation}
\begin{aligned}
\frac{1}{\sigma}\,\frac{\mathrm{d}\sigma^\pm}{\mathrm{d}\cos\theta^*}
&=
\frac{3}{8}
\left[(1\mp \cos\theta^*)^2 f_{\mathrm{L}}^\pm
+ (1\pm \cos\theta^*)^2 f_{\mathrm{R}}^\pm \right. \\
&\qquad\left.
+ 2(1-\cos^2\theta^*) f_{0}^\pm\right],
\end{aligned}
\label{eq:angdist}
\end{equation}
where the upper (lower) sign corresponds to \Wp (\Wm), and $f_{\mathrm{L}}^\pm$, $f_{\mathrm{R}}^\pm$ and $f_0^\pm$ are the left-handed, right-handed and longitudinal helicity fractions, which satisfy $f_{\mathrm{L}}^\pm + f_{\mathrm{R}}^\pm + f_0^\pm = 1$. Higher-order corrections add angular coefficients but retain the helicity-sensitive \cost dependence.

In the presence of an undetected neutrino, $\theta^*$ cannot be reconstructed unambiguously on an event-by-event basis.  However, its effect is reflected in the kinematic distributions of the muon. In particular, for fixed \W boson kinematics, the joint (\ptm , \etam) distribution is sensitive to the helicity state~\cite{Manca:2017xym}, motivating the template approach used here and in Ref.~\cite{CMS:2020cph}.

The dependence of the (\ptm , \etam) distribution on the \W-boson helicity is implemented through a weighting of generator-level events in the $4\pi$ samples because the generators do not directly provide event-by-event helicity information.  For each generator-level $\W\to\mu\nu_\mu$ decay, the $\cos\theta^{*}$ variable is computed in the Collins--Soper frame using the true \W-boson and muon four-momenta. In bins of $\yW$  and $\ptw$, weights used to obtain the left-handed, right-handed and longitudinal distributions are derived by weighting the simulated $\cos\theta^*$ distribution to the analytical form of Eq.~(\ref{eq:angdist}) with the corresponding helicity fraction fixed to unity and others to zero.  These weights transform the original mixture of helicity states produced by the generator into pure left-, right-handed or longitudinal samples, respectively. The weights, computed from the true generator-level \cost distribution, are applied to the corresponding reconstructed event to build, in bins of \yW , two-dimensional templates of \ptm and \etam for pure helicity states. The helicity states populate different regions of the $(\pt^\mu,\eta^\mu)$ plane because the decay angle controls both the direction and transverse momentum of the muon: left- and right-handed bosons emit it preferentially along or against the boson flight direction, shifting it in $\eta^\mu$, while longitudinal bosons emit it transverse to this direction, giving larger $\pt^\mu$. Selected helicity templates illustrating the different kinematic distributions for left- and right-handed \W bosons and their dependence on \yW are shown in the End Matter.

The templates for the main background processes are constructed directly from simulation. The exception is the multijet background from hadrons misidentified as muons, which is estimated from data using a selection with inverted muon-identification criteria. These templates are normalized using theoretical cross-sections, generator efficiencies and correction factors from data where appropriate. 

The helicity fractions are determined from a binned extended maximum-likelihood fit~\cite{JAMES1975343,Dembinski:2022ios,Rodrigues:2020syo} to the (\ptm , \etam) distribution in the fiducial region, implemented with a custom morphed-template cost function. Separate fits are performed for positively and negatively charged muons. The expected number of events in bin $i$ of the (\ptm , \etam) distribution is written as
\begin{align}
\mu_i(\vec{\theta})
&= \sum_{j} \left[
N^{j}_{\mathrm{L}}\, T^{\mathrm{L},j}_{i}(\vec{\theta}_{\mathrm{S}})
+ N^{j}_{\mathrm{R}}\, T^{\mathrm{R},j}_{i}(\vec{\theta}_{\mathrm{S}})
+ N^{j}_{0}\, T^{0,j}_{i}(\vec{\theta}_{\mathrm{S}})
\right]\nonumber \\
&\quad + \sum_{k} N^{(k)}_{\mathrm{bkg}}\, B^{(k)}_{i},
\label{eq:fit}
\end{align}
where the index $j$ runs over the $\yW$ bins, which are fitted simultaneously in a single likelihood for each charge. The full set of parameters entering the expected bin contents is denoted by $\vec{\theta}$: the yields $N^{j}_{\mathrm{L,R,0}}$ of the left-handed, right-handed and longitudinal signal components in each $\yW$ bin, the background yields $N^{(k)}_{\mathrm{bkg}}$, and the QCD-scale shape nuisance parameters, collectively denoted $\vec{\theta}_{\mathrm{S}}$. The quantities $T^{\lambda,j}_{i}$, with $\lambda = \mathrm{L}, \mathrm{R}, 0$, are the signal templates normalized to unit integral, \ie the fraction of the corresponding template found in bin $i$, such that the $N$ parameters are pure yields; $B^{(k)}_{i}$ is defined analogously for background component $k$. The shape nuisance parameters modify the signal templates through a linear morphing, $T^{\lambda,j}_{i}(\vec{\theta}_{\mathrm{S}}) \propto \tilde{T}^{\lambda,j}_{i} + \sum_{m} \theta_{m} \Delta^{\lambda,j}_{i,m}$, where $\tilde{T}^{\lambda,j}_{i}$ is the nominal template and $\Delta^{\lambda,j}_{i,m}$ is the shape variation associated with nuisance parameter $\theta_{m}$. 
The likelihood is the product over bins of Poisson probabilities for the observed event counts given the expectations $\mu_{i}(\vec{\theta})$, multiplied by normally distributed terms for the shape nuisance parameters described below.
The normalizations of all background templates, except for the hadronic mis-ID background, are fixed to their expectations from simulation, with the associated cross-section and modeling uncertainties assessed through the systematic variations described below. The normalization of the hadronic mis-ID template is a free parameter in the fit, obtained from the data. Given the limited sensitivity to the longitudinal component in the forward acceptance, the longitudinal yields $N^{j}_{0}$ are fixed to the SM expectation~\cite{Alwall:2014hca}. 
The left- and right-handed yields $N^{j}_{\mathrm{L}}$ and $N^{j}_{\mathrm{R}}$ are free parameters of the fit, so that the total transverse signal yield in each \yW bin is determined from the data. 
The helicity fractions within the fiducial acceptance in each $\yW$ bin are obtained from the fitted yields as $f^{j}_{\lambda} = N^{j}_{\lambda} / (N^{j}_{\mathrm{L}} + N^{j}_{\mathrm{R}} + N^{j}_{0})$. 

While signal templates are generated in bins of \yW , the fit is additionally sensitive to the \ptw shape agreement between data and simulation. The renormalization, $\mu_{\mathrm{R}}$, and factorization, $\mu_{\mathrm{F}}$, scale uncertainties are therefore incorporated directly in the fit as Gaussian-constrained shape nuisance parameters, constructed from six independent variations of the two scales by factors of two~\cite{Hamilton:2013fea}. The six scale-variation templates are evaluated in bins of \ptw, and their per-bin shape differences with respect to the nominal template enter the likelihood as Gaussian-constrained nuisance parameters. Variations whose signal-to-noise ratio falls below threshold are discarded, as they are indistinguishable from the statistical noise of the simulated sample and would otherwise introduce parameters that fit simulation statistical fluctuations rather than physics effects. Statistical uncertainties of the templates, including those arising from the helicity weighting, are treated with a Barlow--Beeston-like method~\cite{Dembinski:2022ios}, in which each template bin is a Gaussian-constrained nuisance parameter around its nominal expectation with variance given by the simulation statistical uncertainty. The penalized likelihood is maximized using the \mbox{\textsc{iMinuit}} minimizer~\cite{py:iminuit} and parameter uncertainties are derived from the profile likelihood. 

The fitting procedure is validated with closure tests on pseudoexperiments~\cite{Cowan:2010js}. Pseudodata samples are constructed by weighting simulated events to specific helicity fractions, including values differing from the SM prediction. The full template construction, fit configuration and systematic treatment are applied in these tests, and the extracted fractions are compared with their input values. No significant biases are observed.

Experimental uncertainties arise from the muon momentum scale and resolution, the track-reconstruction, muon-identification and trigger efficiencies, the modeling of the isolation and impact-parameter distributions, and the \ptw correction. Each source is varied within its measured uncertainty: the momentum calibration within its covariance, the efficiency scale factors within their uncertainties, the isolation and impact-parameter modeling by adjusting thresholds and weighting to alternative control samples, and the \ptw correction by propagating the statistical and smoothing uncertainties of the data-to-simulation ratio. The template construction and fit are repeated for each. Background uncertainties cover the cross-sections and kinematic distributions of the simulated backgrounds, and the shape and normalization of the hadronic mis-ID background, with the latter assessed using alternative parameterizations and selections on the control sample. In each \yW bin the corresponding variation of the extracted helicity fractions is taken as the uncertainty.

Theoretical and modeling uncertainties affecting the signal templates are evaluated using variations on the generator-level samples. The renormalization and factorization scale uncertainties are extracted from the nuisance parameters' contribution to the post-fit covariance matrix. The uncertainties due to the choice of PDF are evaluated by comparing the results obtained using replica sets of the nominal \textsc{NNPDF2.3} set~\cite{Ball:2012cx}.  For each PDF set replica, new helicity weights, signal templates, and scale variations are constructed and the full fit is repeated. The standard deviation of the replica results with respect to the nominal extraction is taken as the systematic uncertainty. An additional uncertainty accounts for the helicity-dependent transfer from the boson rest frame decay angle to the reconstructed muon kinematics within the fiducial acceptance. Since the fiducial requirements select different regions of \cost phase space for different helicity states, this transfer depends on the modeling of the \W production and decay. The effect is estimated from the difference between the nominal \MG prediction and alternative \powheg predictions for the \yW-dependent fiducial \cost distribution, taken as a regularized two-point envelope.

The systematic contributions not already profiled in the fit are assumed to be uncorrelated and are combined in quadrature for each \yW bin and each helicity fraction, while the renormalization and factorization scale variations are propagated directly through the fit.

The \yW binning is optimized separately for \Wp and \Wm bosons based on the expected yields across this distribution. For each bin, $f_{\textrm{L}}$ and $f_{\textrm{R}}$ are extracted in the fiducial region defined by $2.2 \leq \etam \leq 4.4$ and $27\leq \ptm \leq 45\gev$, while $f_{\textrm{0}}$ is estimated as described above. At large \yW the total uncertainty has a significant contribution from the size of the data sample, while the leading systematic source varies by bin: the fiducial helicity transfer dominates at low \yW and the PDF uncertainty at high \yW.

\begin{figure}[!hbpt]
\begin{center}
  \includegraphics[trim= 0mm 0mm 0mm 0mm ,clip, width=0.45\textwidth]{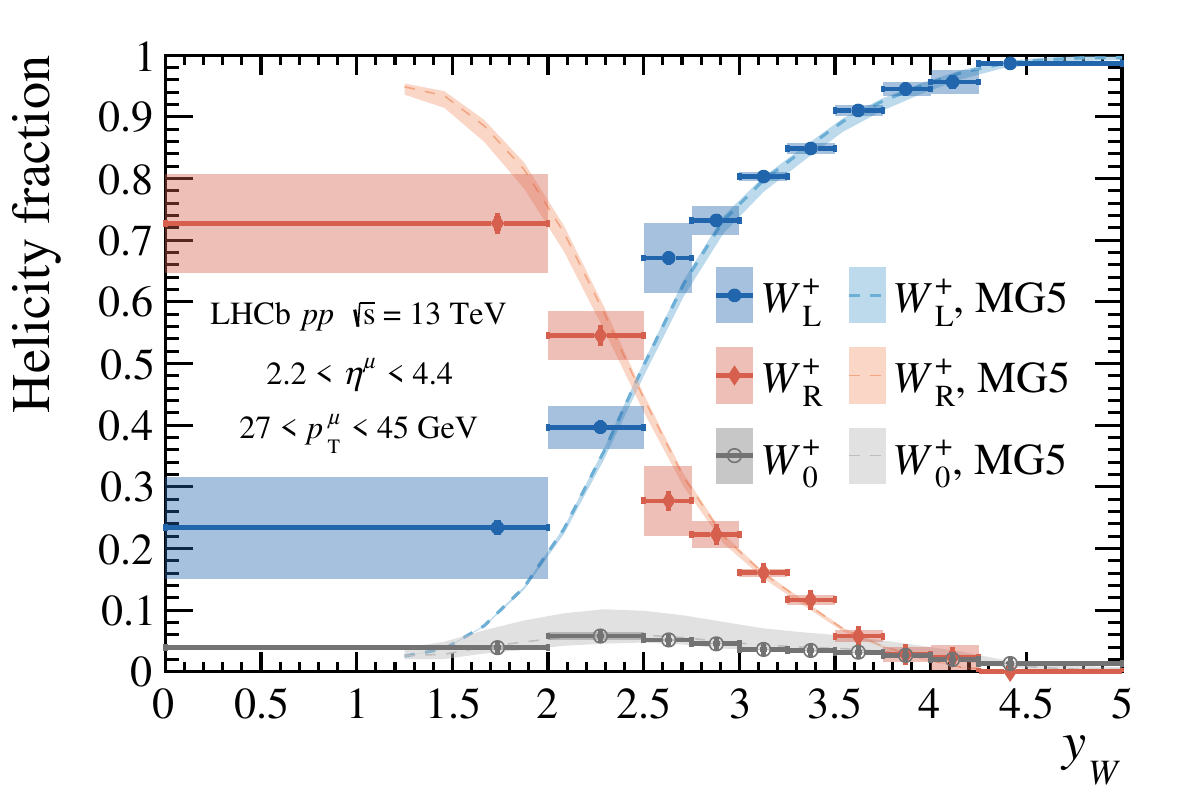}
  \includegraphics[trim= 0mm 0mm 0mm 0mm ,clip, width=0.45\textwidth]{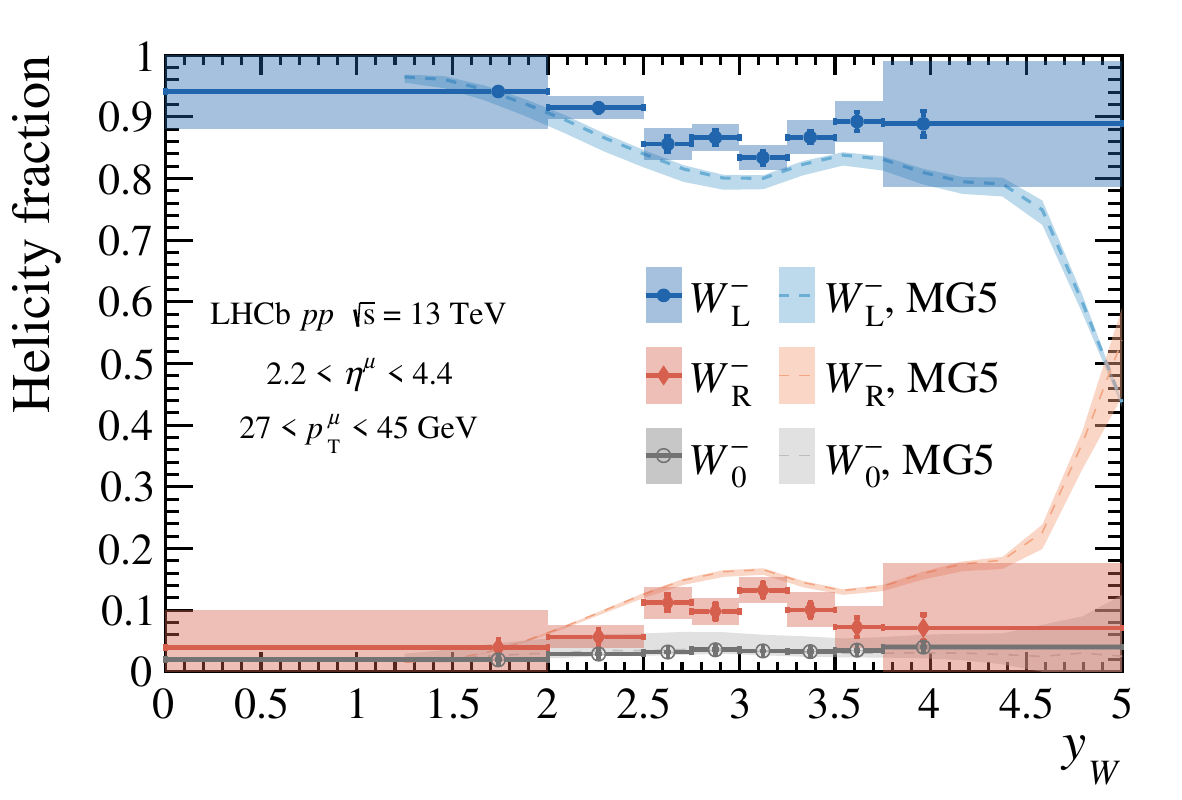} \\

\caption{Results for $f_{\textrm{L}}$ and $f_{\textrm{R}}$ of (left) \Wp and (right) \Wm  bosons as a function of \yW.  Points and vertical bars show the fitted values and uncertainties, and are placed at the expected mean \yW in each bin. The filled blocks correspond to the systematic uncertainties. The dashed curves show the \MG predictions, MG5, with bands corresponding to the PDF and scale uncertainties, as defined in the text. The prediction for the longitudinal component, whose yield is fixed in the fit, is shown for reference. }
\label{fig:results}
    \end{center}
\end{figure}

The measured transverse helicity fractions are presented as a function of \yW in Fig.~\ref{fig:results}, and align closely with the expectation illustrated in the Supplemental Material~\cite{SupplementalMaterial}. For \Wp bosons, the left-handed fraction $f_{\textrm{L}}^+$ rises from $0.23$ in the lowest \yW bin to $0.99$ at the highest \yW, while for \Wm bosons, $f_{\textrm{L}}^-$ varies between $0.83$ and $0.94$ across \yW. The results are compared with NLO predictions obtained using \MG ~\cite{Alwall:2014hca} and several modern PDF sets~\cite{NNPDF:2017mvq,Hou:2019efy,Bailey:2020ooq}. Within the total experimental and theoretical uncertainties, the measured transverse helicity fractions are consistent with the SM expectation across the full \yW range; however, a coherent trend appears for both \Wp and \Wm bosons, with the left-handed fractions above (and right-handed fractions below) the SM prediction at the level of approximately one standard deviation in the individual \yW intervals. 

In summary, this Letter extends \W boson helicity measurements to the forward region, where asymmetric partonic kinematics produce a particularly pronounced separation of the two transverse states. The results test a characteristic consequence of the chiral charged-current interaction in a previously unexplored phase space and confirm the expected rapidity dependence of the helicity composition. The numerical results are available in HEPData~\cite{HepData}. The measurement provides a new benchmark for the angular and fiducial modeling of forward charged-current production, of direct relevance to future precision electroweak and PDF-sensitive measurements at the LHC.


\section*{Acknowledgements}
%
%
\noindent We express our gratitude to our colleagues in the CERN
accelerator departments for the excellent performance of the LHC. We
thank the technical and administrative staff at the LHCb
institutes.
We acknowledge support from CERN and from the national agencies:
ARC (Australia);
CAPES, CNPq, FAPERJ and FINEP (Brazil); 
MOST and NSFC (China); 
CNRS/IN2P3 and CEA (France);  
BMFTR, DFG and MPG (Germany);
NKFIH (Hungary);              
INFN (Italy); 
NWO (Netherlands); 
MNiSW and NCN (Poland); 
MEC/IFA (Romania); 
MICIU and AEI (Spain);
SNSF and SER (Switzerland); 
NASU (Ukraine); 
STFC (United Kingdom); 
DOE NP and NSF (USA).
We acknowledge the computing resources that are provided by ARDC (Australia), 
CBPF (Brazil),
CERN, 
IHEP and LZU (China),
IN2P3 (France), 
KIT and DESY (Germany), 
INFN (Italy), 
SURF (Netherlands),
Polish WLCG (Poland),
IFIN-HH (Romania), 
PIC (Spain), CSCS (Switzerland), 
GridPP (United Kingdom),
and NSF (USA).  
We are indebted to the communities behind the multiple open-source
software packages on which we depend.
Individual groups or members have received support from
RTP (Australia), 
FWO Odysseus grant G0ASD25N (Belgium), 
Key Research Program of Frontier Sciences of CAS, CAS PIFI, CAS CCEPP (China); 
Minciencias (Colombia);
EPLANET, Marie Sk\l{}odowska-Curie Actions, ERC and NextGenerationEU (European Union);
A*MIDEX, ANR, IPhU and Labex P2IO, and R\'{e}gion Auvergne-Rh\^{o}ne-Alpes (France);
Alexander-von-Humboldt Foundation (Germany);
ICSC (Italy); 
Severo Ochoa and Mar\'ia de Maeztu Units of Excellence, GVA, XuntaGal, GENCAT, InTalent-Inditex and Prog.~Atracci\'on Talento CM (Spain);
the Leverhulme Trust, the Royal Society and UKRI (United Kingdom).

\clearpage

\section*{End Matter}

\subsection*{Helicity templates}

\begin{figure}[!b]
\begin{center}
  \includegraphics[trim= 0mm 0mm 0mm 0mm ,clip, width=0.45\textwidth]{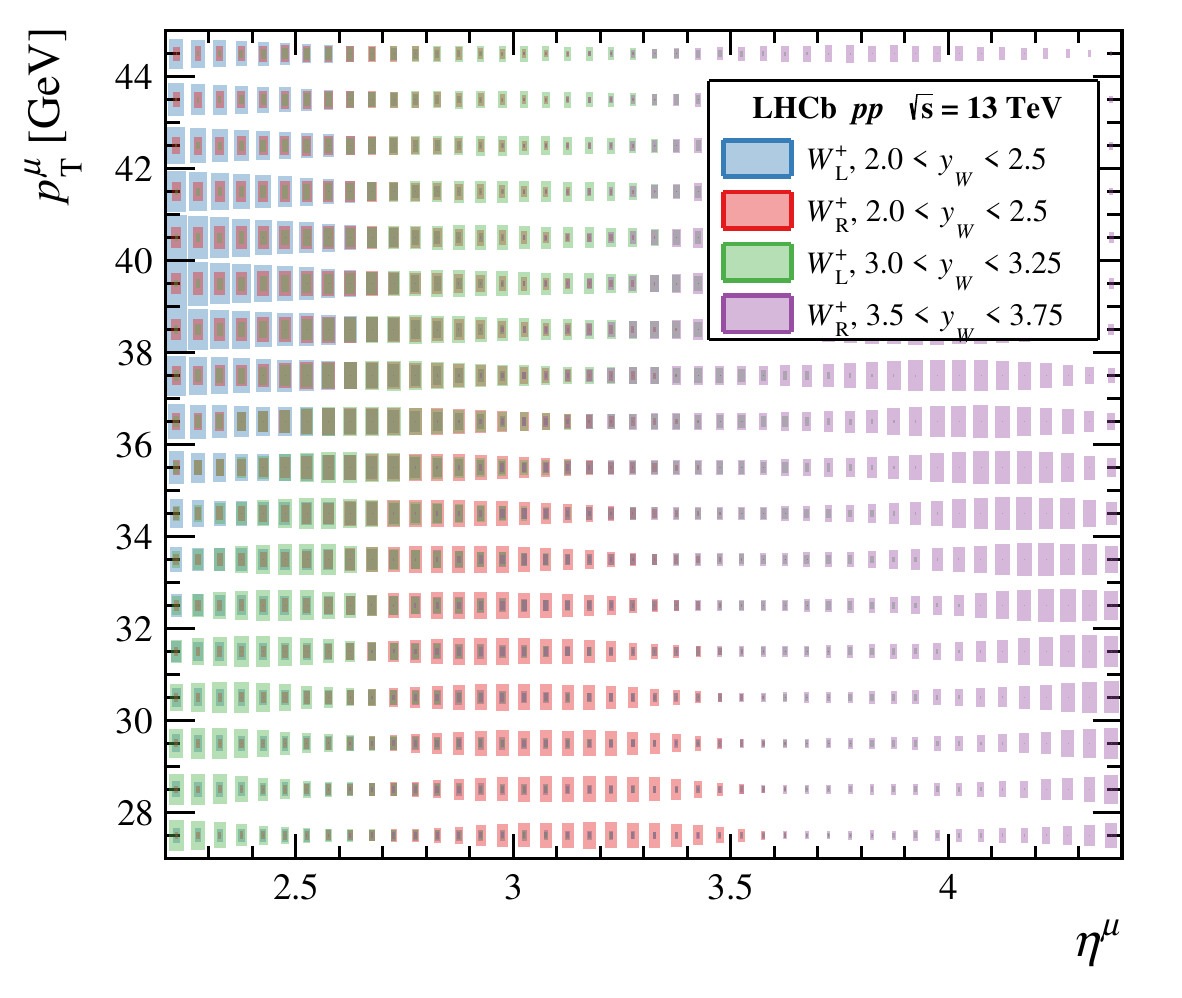}
  \includegraphics[trim= 0mm 0mm 0mm 0mm ,clip, width=0.45\textwidth]{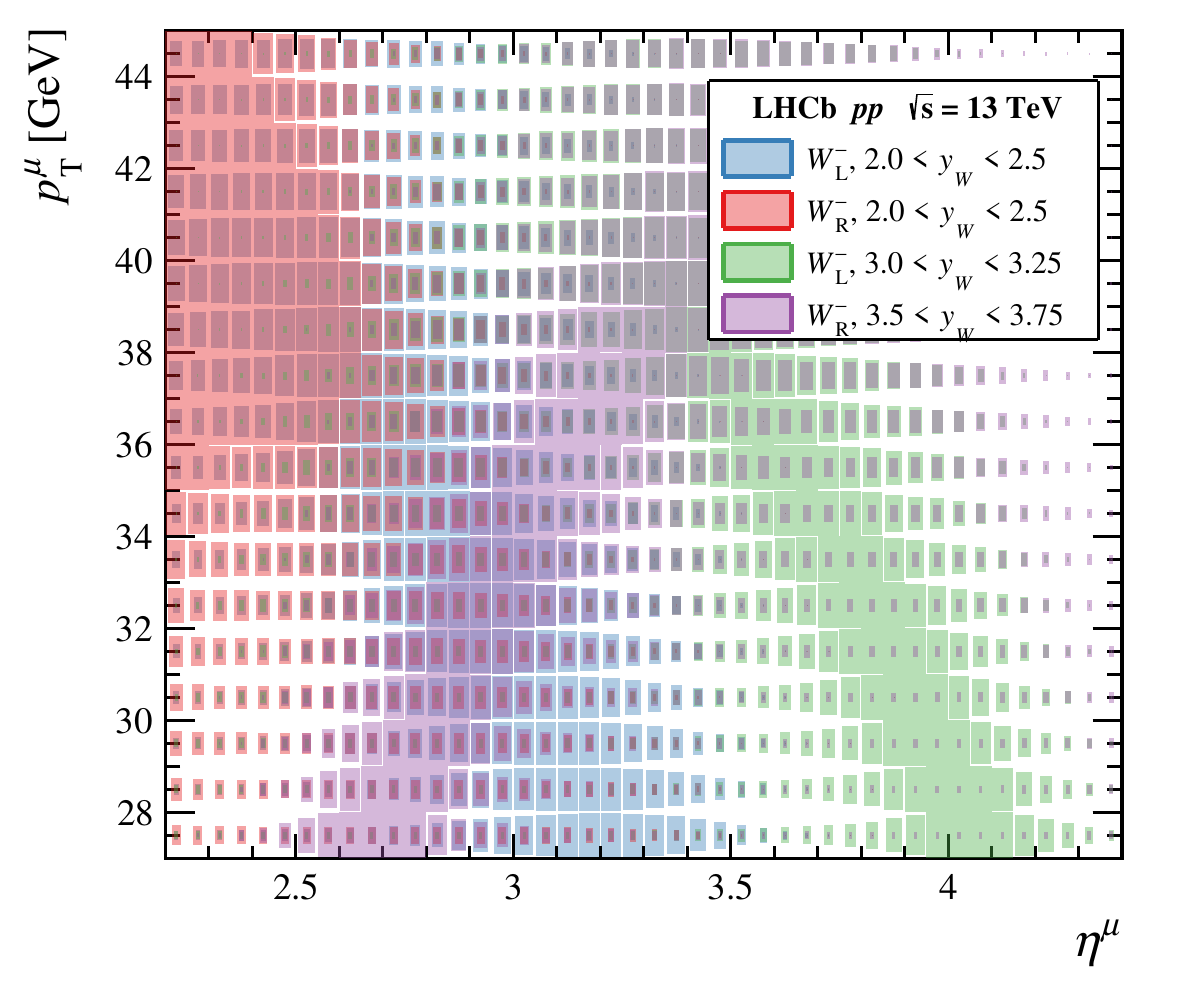} \\

\caption{Reconstructed templates of $\etam$ and $\ptm$ for (left) \Wp  and (right) \Wm  decays for left-handed and right-handed states in different \yW bins.  The concentration of the colors corresponds to normalized yields. The difference between the \Wp and \Wm templates reflects the charge-asymmetric parton content of the proton.}
\label{fig:TemplatesAll}
    \end{center}
\end{figure}

The helicity weights are applied to fully reconstructed simulated events to build, in each \yW bin, two-dimensional templates of \ptm and \etam corresponding to pure left-handed, right-handed and longitudinal \W bosons. A representative selection of templates is shown in Fig.~\ref{fig:TemplatesAll} for both charges. The left- and right-handed templates populate different regions of the  (\ptm, \etam) plane, and the difference between them evolves with \yW. This shape separation is what allows the transverse fractions to be extracted from an observable that does not contain \cost directly. The difference between the \Wp and \Wm panels reflects the charge-asymmetric valence content of the proton rather than any detector effect.

\subsection*{Post-fit distributions}
 
\begin{figure}[!hbpt]
\begin{center}
  \includegraphics[trim= 0mm 0mm 0mm 0mm ,clip, width=0.45\textwidth]{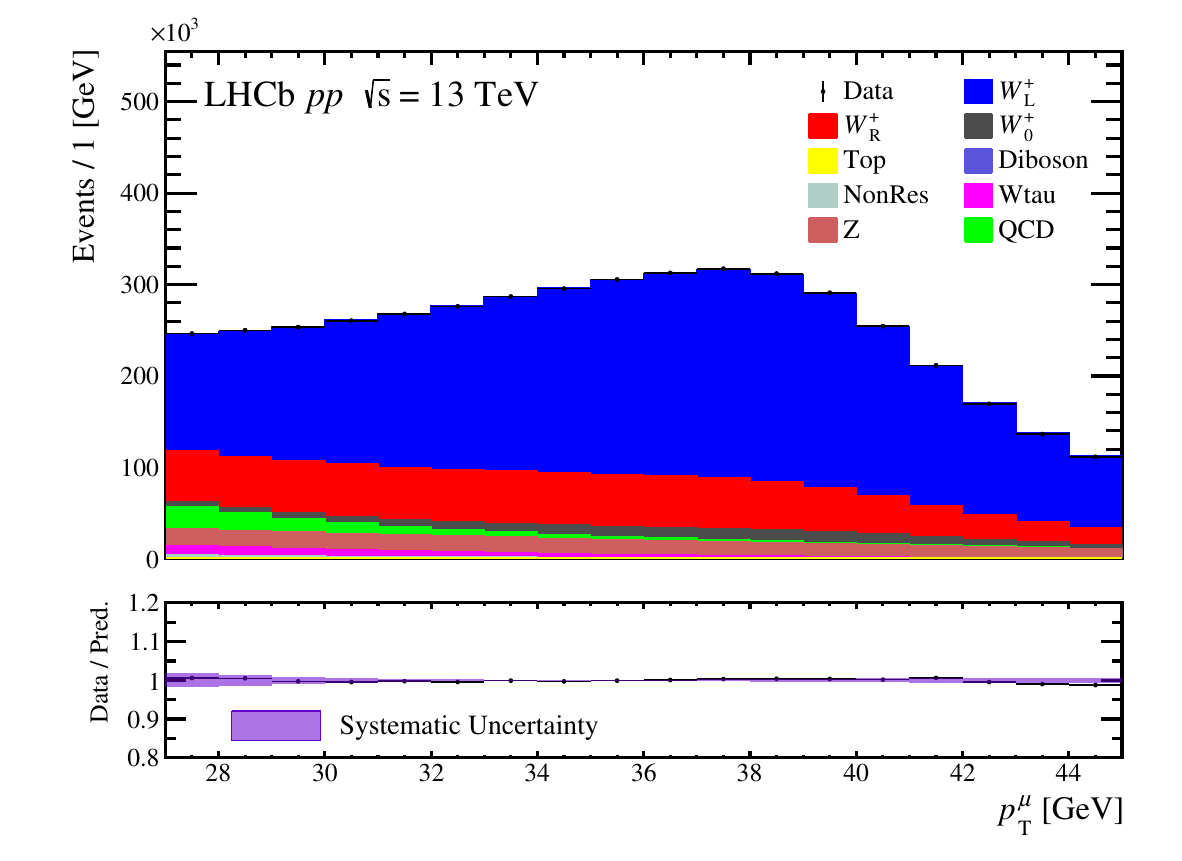}
  \includegraphics[trim= 0mm 0mm 0mm 0mm ,clip, width=0.45\textwidth]{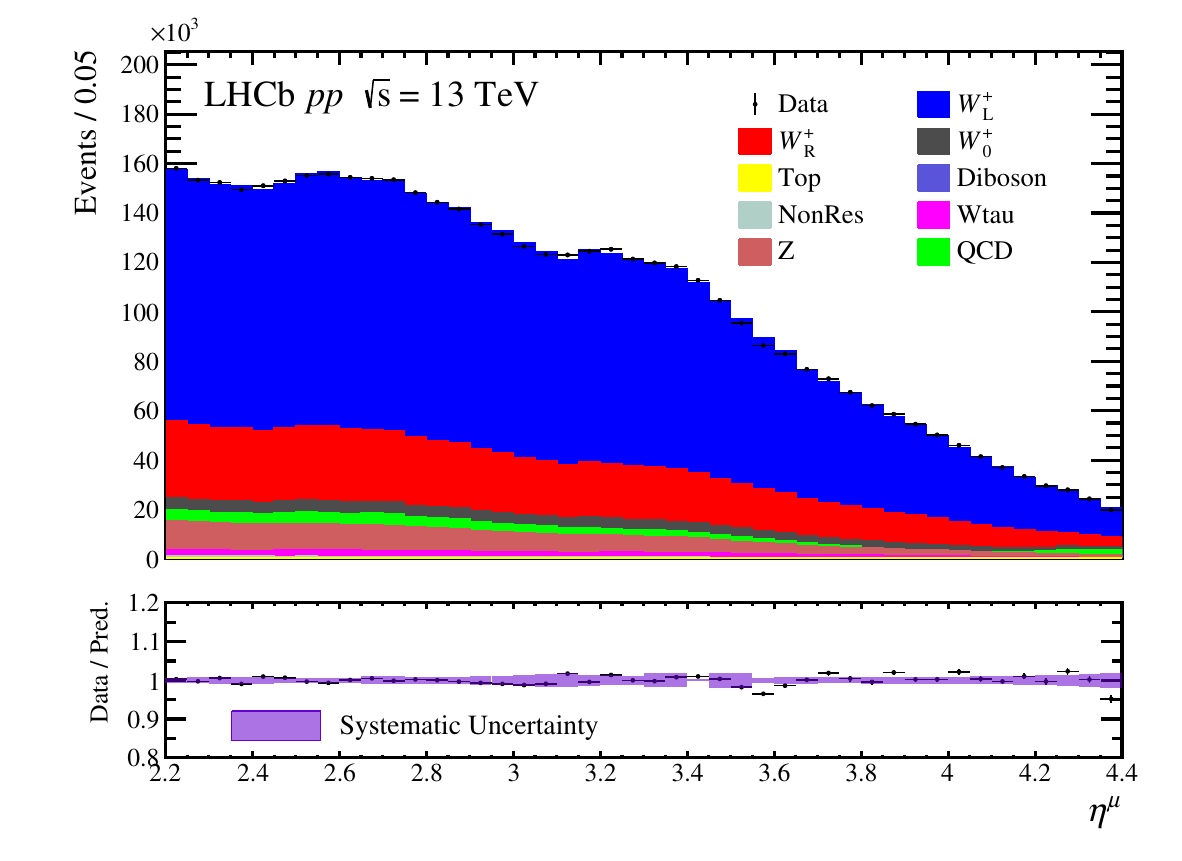} \\
  \includegraphics[trim= 0mm 0mm 0mm 0mm ,clip, width=0.45\textwidth]{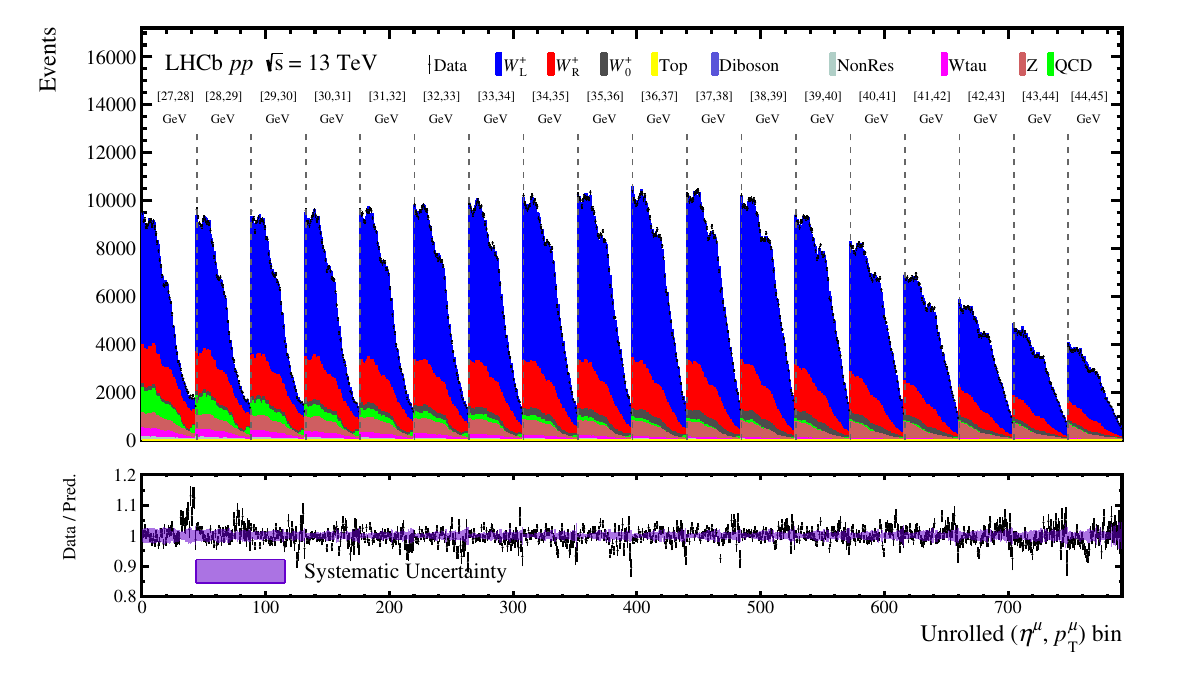}

\caption{Post-fit distributions of 1-dimensional projections of the (top left) \ptm, (top right) \etam, and (bottom) the full unrolled fit distribution for \Wp helicity states.  Each state has the contribution of each \yW bin combined into a single contribution on the plot.  Each tooth of the saw-tooth pattern corresponds to a \ptm bin. The purple band in the ratio plot corresponds to the systematic uncertainty coverage of the considered kinematic variable.}
\label{fig:PostFitWp}
    \end{center}
\end{figure}

\begin{figure}[!hbpt]
\begin{center}
  \includegraphics[trim= 0mm 0mm 0mm 0mm ,clip, width=0.45\textwidth]{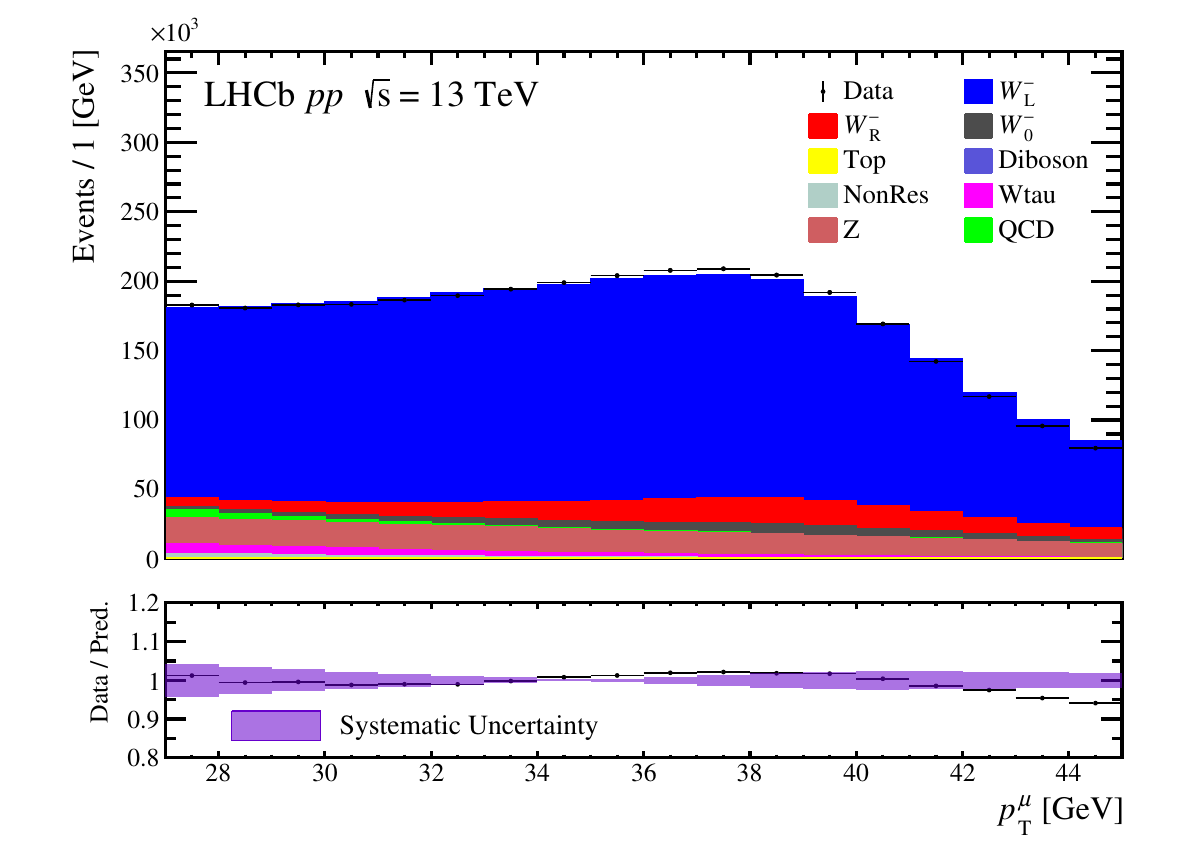}
  \includegraphics[trim= 0mm 0mm 0mm 0mm ,clip, width=0.45\textwidth]{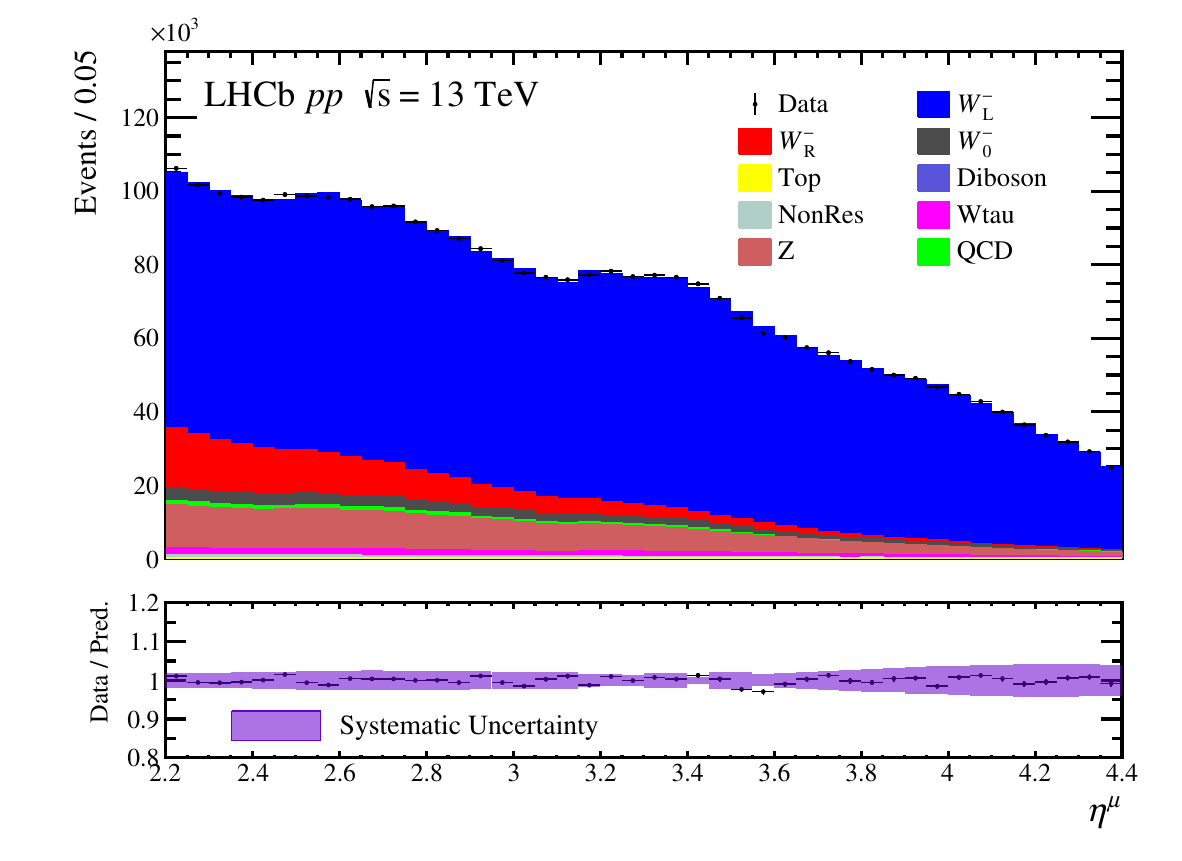} \\
  \includegraphics[trim= 0mm 0mm 0mm 0mm ,clip, width=0.45\textwidth]{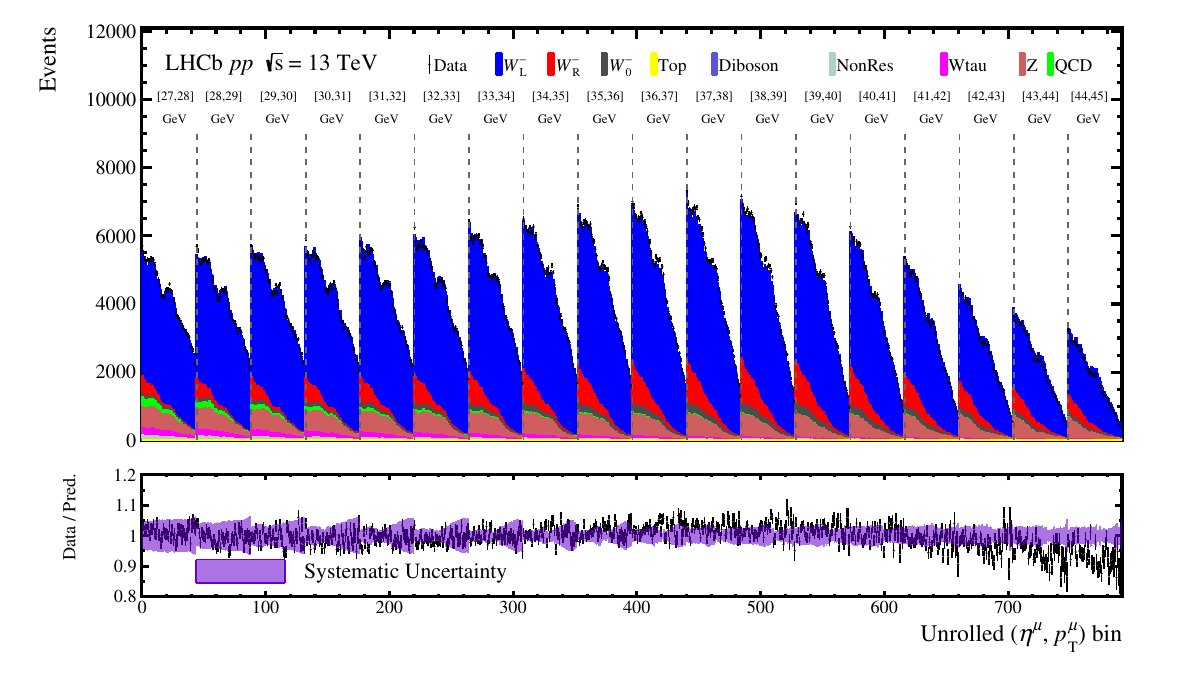}

\caption{Post-fit distributions of 1-dimensional projections of the (top left) \ptm, (top right) \etam, and (bottom) the full unrolled fit distribution for \Wm helicity states.  Each state has the contribution of each \yW bin combined into a single contribution on the plot.  Each tooth of the saw-tooth pattern corresponds to a \ptm bin.  The purple band in the ratio plot corresponds to the systematic uncertainty coverage of the considered kinematic variable.}
\label{fig:PostFitWm}
    \end{center}
\end{figure}

Figures~\ref{fig:PostFitWp} and~\ref{fig:PostFitWm} show the post-fit one-dimensional projections in $\pt^{\mu}$ and $\eta^{\mu}$, together with the full unrolled distribution, for \Wp and \Wm bosons respectively, with the contributions of the individual \yW bins combined. Each tooth of the saw-tooth pattern corresponds to one \ptm bin. The description of the data is good across the fitted range. In the highest \ptm bins the data lie below the post-fit prediction; these are the least populated bins of the spectrum, and the shift is compatible with the statistical uncertainty of the data. Additional formatting of these distributions is available in the Supplemental Material~\cite{SupplementalMaterial}.

\subsection*{Systematic Uncertainties}

Table ~\ref{tab:wheli_syst_ranges} shows the statistical and systematic uncertainty ranges per \yW interval, summed over source subcategories.

\begin{table}[!htbp]
  \centering
  \caption{Range of the absolute uncertainty on each helicity fraction, by source, over all \yW bins, for \Wp and \Wm. Each entry gives the smallest and the largest value across the bins. The total is the quadrature sum of the statistical uncertainty and all systematic contributions. Values are quoted to two significant figures.}
  \label{tab:wheli_syst_ranges}
    \resizebox{\textwidth}{!}{\begin{tabular}{l r@{ -- }lr@{ -- }lr@{ -- }l:r@{ -- }lr@{ -- }lr@{ -- }l}
    \hline\hline
     & \multicolumn{6}{c:}{\Wp} & \multicolumn{6}{c}{\Wm} \\
    \cline{2-13}
    Source & \multicolumn{2}{c}{$f_{L}$} & \multicolumn{2}{c}{$f_{R}$} & \multicolumn{2}{c:}{$f_{0}$} & \multicolumn{2}{c}{$f_{L}$} & \multicolumn{2}{c}{$f_{R}$} & \multicolumn{2}{c}{$f_{0}$} \\
    \hline
    Smearing & $9.6\times10^{-5}$ & $0.014$ & $1.1\times10^{-6}$ & $0.013$ & $7.6\times10^{-5}$ & $0.00084$ & $0.0033$ & $0.039$ & $0.0034$ & $0.040$ & $0.00022$ & $0.0016$ \\
    Efficiencies & $0.00051$ & $0.031$ & $5.9\times10^{-7}$ & $0.032$ & $5.8\times10^{-5}$ & $0.00081$ & $0.0079$ & $0.054$ & $0.0078$ & $0.056$ & $9.7\times10^{-5}$ & $0.0017$ \\
    QCD background & $7.4\times10^{-5}$ & $0.0036$ & $5.1\times10^{-7}$ & $0.0038$ & $1.5\times10^{-5}$ & $0.00021$ & $0.00075$ & $0.0078$ & $0.00072$ & $0.0080$ & $3.1\times10^{-5}$ & $0.00026$ \\
    PDF & $0.00022$ & $0.0044$ & $2.8\times10^{-7}$ & $0.0048$ & $0.00015$ & $0.00049$ & $0.0017$ & $0.015$ & $0.0015$ & $0.016$ & $0.00021$ & $0.0016$ \\
    Helicity modelling & $0.00090$ & $0.075$ & $1.2\times10^{-6}$ & $0.072$ & $0.00026$ & $0.0058$ & $0.0070$ & $0.071$ & $0.0067$ & $0.075$ & $0.00017$ & $0.0031$ \\
    Scale variations & $6.8\times10^{-5}$ & $0.0077$ & $4.3\times10^{-9}$ & $0.0076$ & $2.2\times10^{-5}$ & $0.00022$ & $0.00031$ & $0.012$ & $0.00023$ & $0.012$ & $2.9\times10^{-5}$ & $0.00057$ \\
    Statistical & $0.00017$ & $0.0081$ & $2.6\times10^{-6}$ & $0.0080$ & $0.00015$ & $0.00060$ & $0.0043$ & $0.021$ & $0.0043$ & $0.021$ & $0.00014$ & $0.00085$ \\
    \hdashline
    Total & $0.0011$ & $0.083$ & $3.1\times10^{-6}$ & $0.081$ & $0.00041$ & $0.0059$ & $0.019$ & $0.10$ & $0.019$ & $0.11$ & $0.00054$ & $0.0043$ \\
    \hline\hline
  \end{tabular}}
\end{table}

\clearpage

\section*{Supplemental material for First measurement of the forward rapidity dependence of $W$ boson transverse helicity fractions }
\label{sec:Supplemental-App}

\begin{figure}[!hbpt]
\begin{center}
  \includegraphics[trim= 0mm 0mm 0mm 0mm ,clip, width=0.45\textwidth]{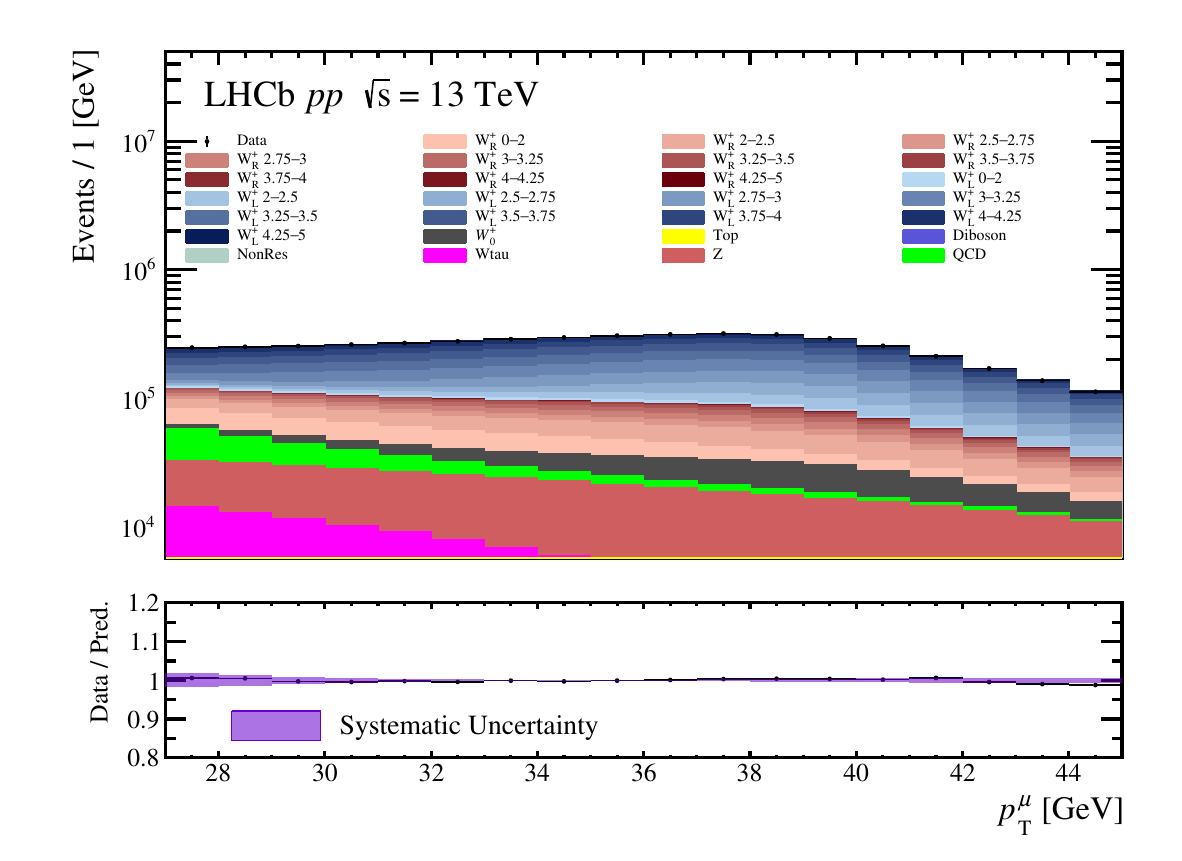}
  \includegraphics[trim= 0mm 0mm 0mm 0mm ,clip, width=0.45\textwidth]{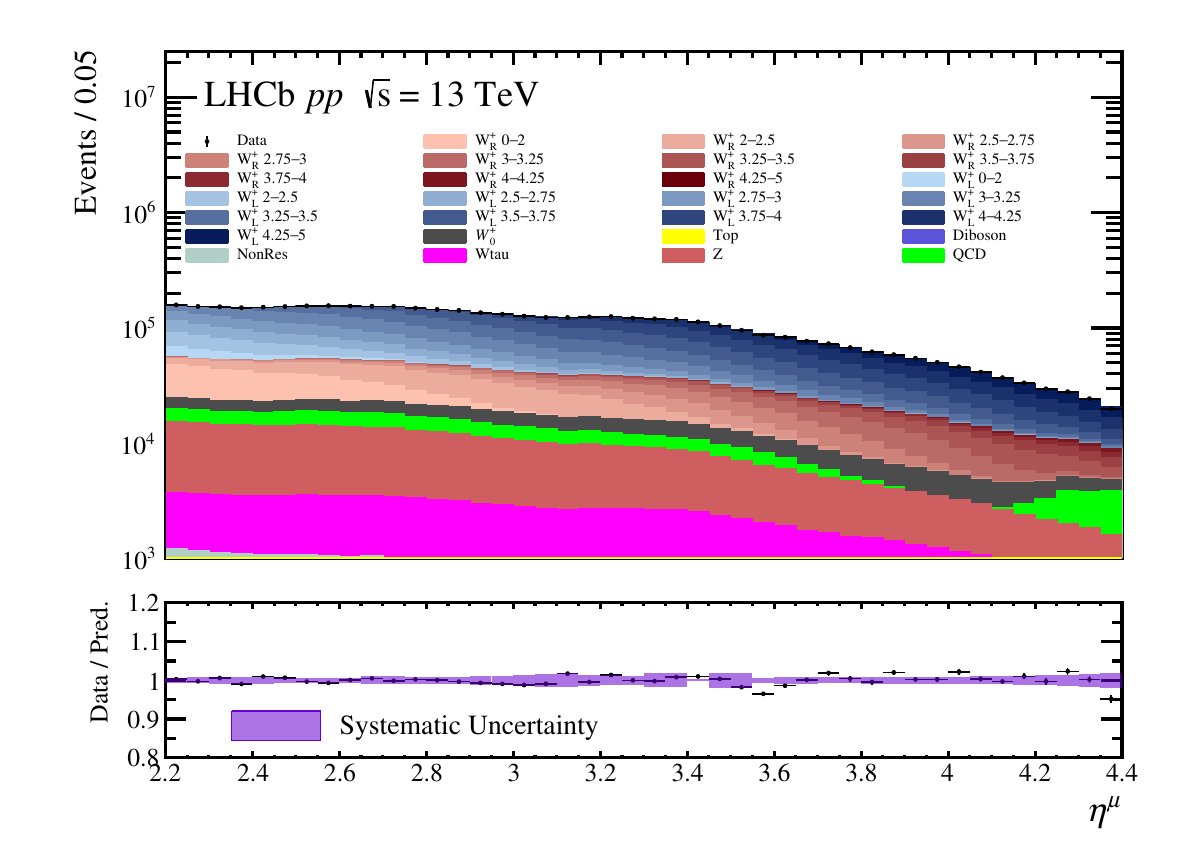} \\
  \includegraphics[trim= 0mm 0mm 0mm 0mm ,clip, width=0.45\textwidth]{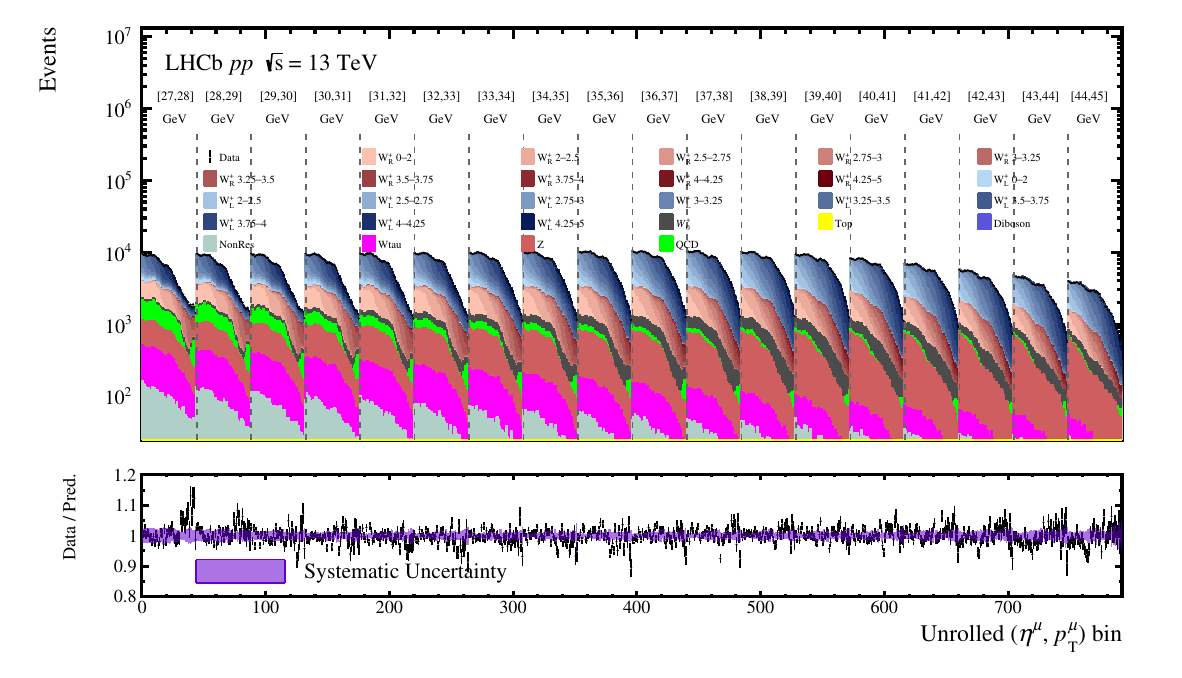}

\caption{Post-fit distributions of 1-dimensional projections of the (top left) \ptm, (top right) \etam, and (bottom) the full unrolled fit distribution for \Wp helicity states.  Each tooth of the saw-tooth pattern corresponds to a \ptm bin. The purple band in the ratio plot corresponds to the systematic uncertainty coverage of the considered kinematic variable.}
\label{fig:PostFitWpUnmerged}
    \end{center}
\end{figure}

\begin{figure}[!hbpt]
\begin{center}
  \includegraphics[trim= 0mm 0mm 0mm 0mm ,clip, width=0.45\textwidth]{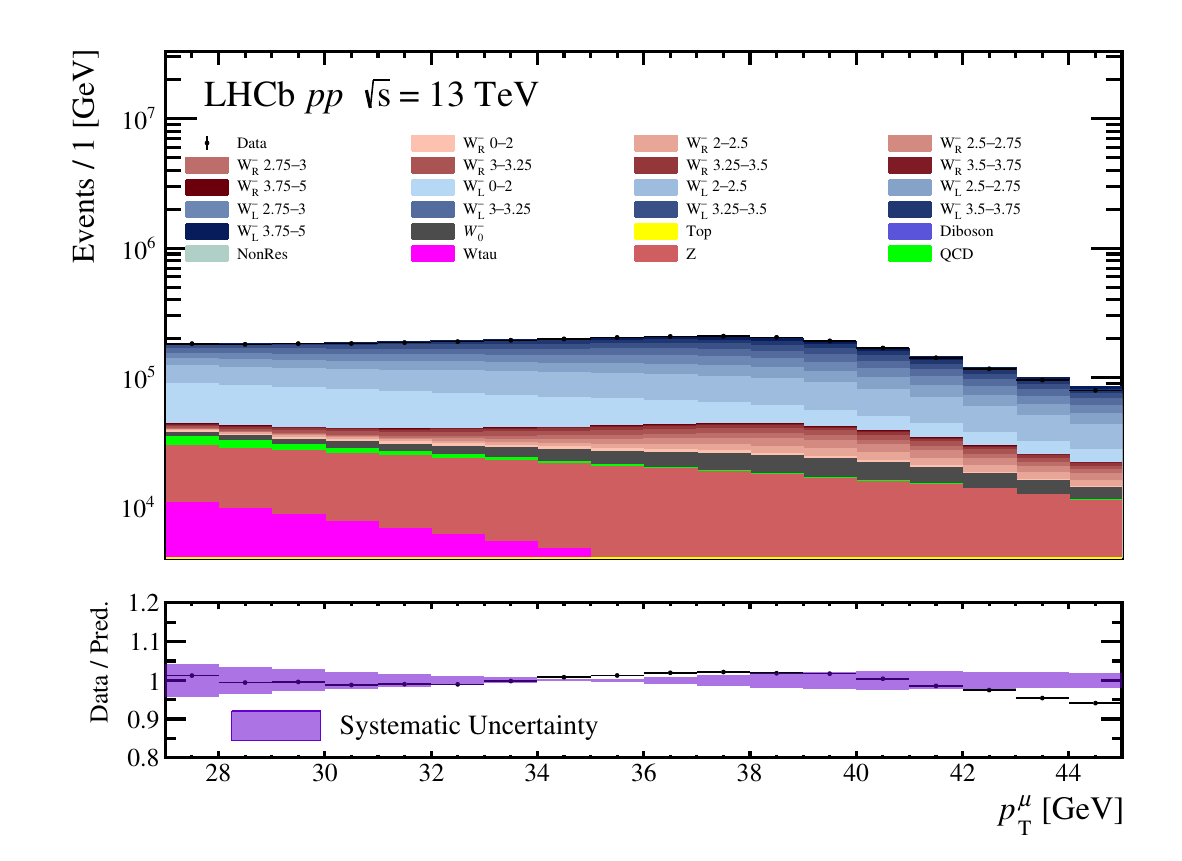}
  \includegraphics[trim= 0mm 0mm 0mm 0mm ,clip, width=0.45\textwidth]{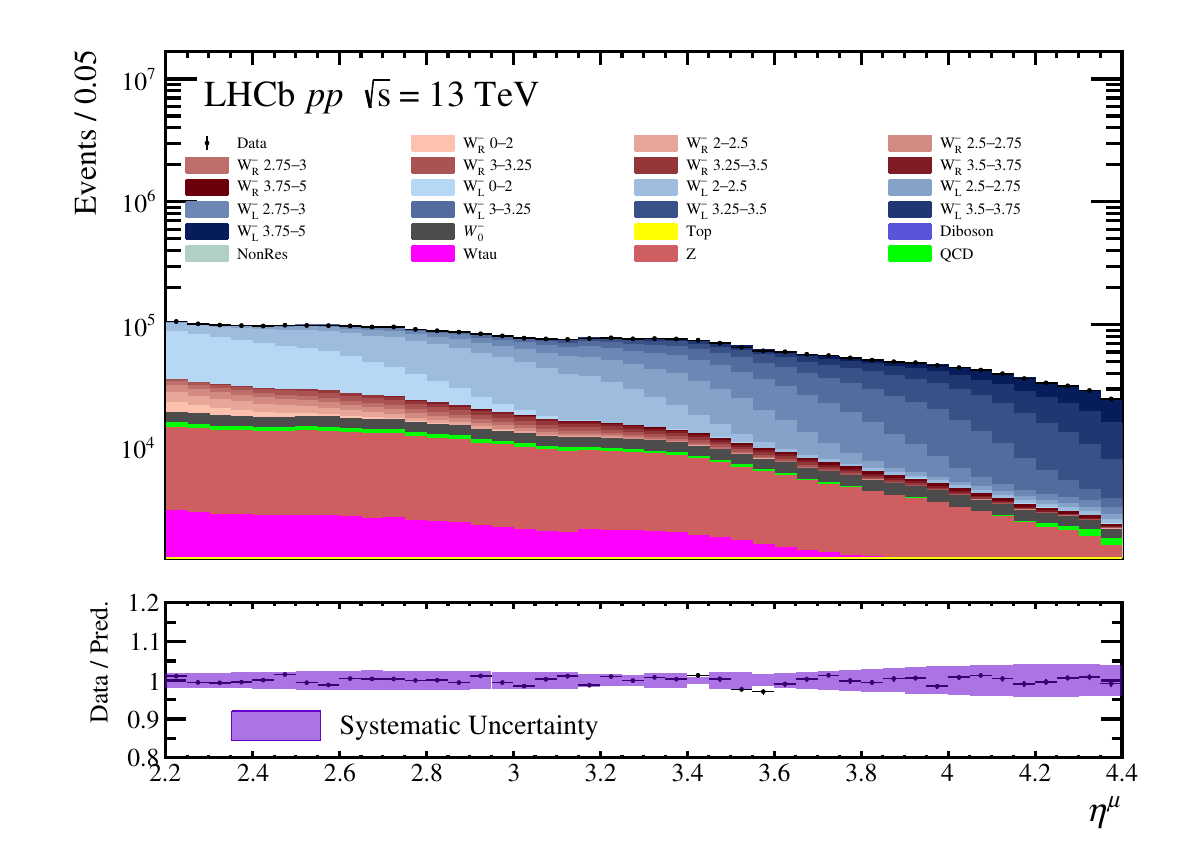} \\
  \includegraphics[trim= 0mm 0mm 0mm 0mm ,clip, width=0.45\textwidth]{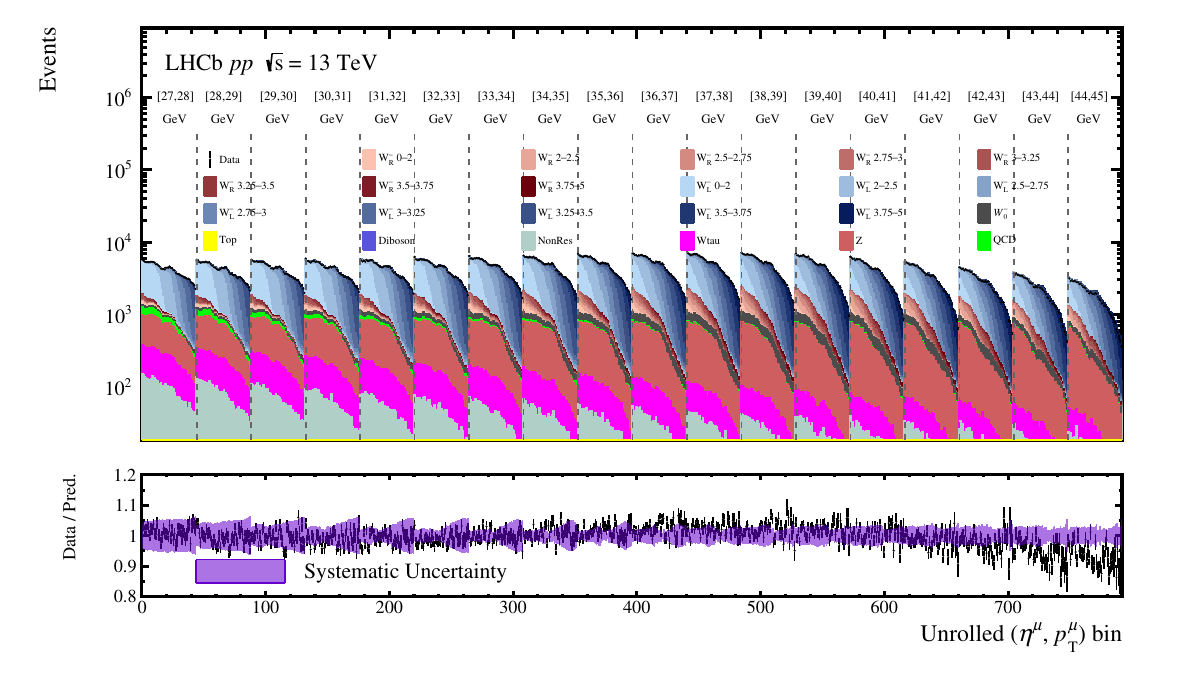}

\caption{Post-fit distributions of 1-dimensional projections of the (top left) \ptm, (top right) \etam, and (bottom) the full unrolled fit distribution for \Wm helicity states.  Each tooth of the saw-tooth pattern corresponds to a \ptm bin.  The purple band in the ratio plot corresponds to the systematic uncertainty coverage of the considered kinematic variable.}
\label{fig:PostFitWmUnmerged}
    \end{center}
\end{figure}

\begin{figure}[!hbpt]
\begin{center}
  \includegraphics[trim= 0mm 0mm 0mm 0mm ,clip, width=0.45\textwidth]{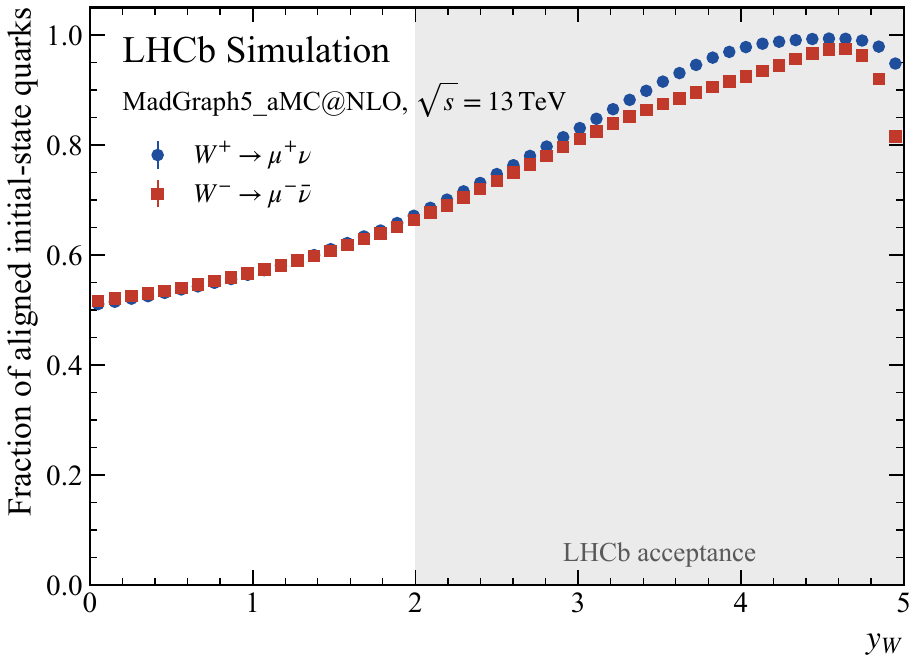}
  \includegraphics[trim= 0mm 0mm 0mm 0mm ,clip, width=0.45\textwidth]{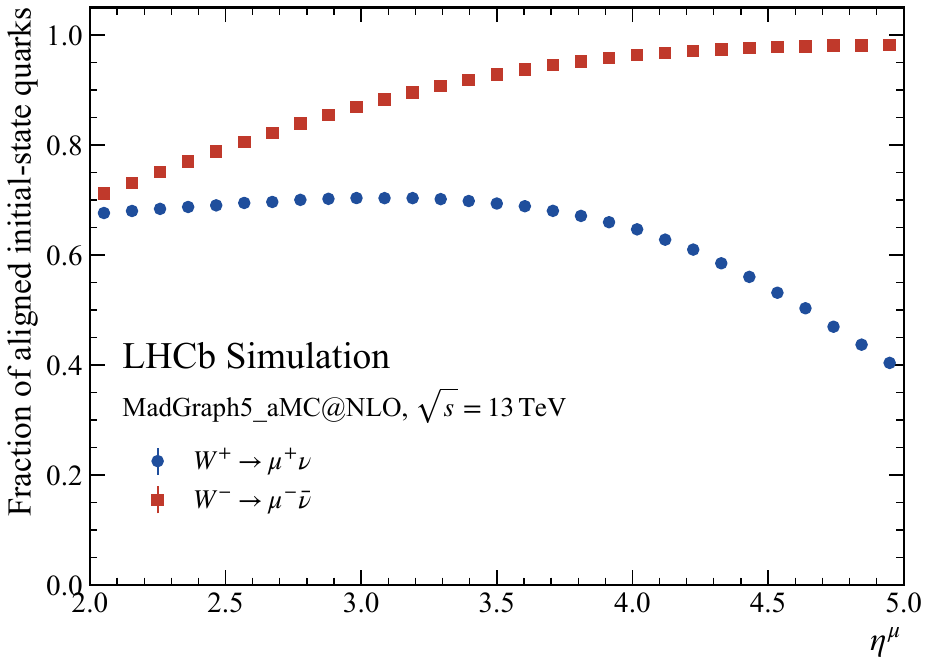} \\

\caption{Fraction of simulated $W^{+}$ (blue circles) and $W^{-}$ (red squares)
    events in which the initial-state quark, rather than the antiquark, moves in the
    same longitudinal direction as the $W$ boson, shown (left) as a function of the
    $W$-boson rapidity, $y_{W}$, and (right) as a function of the pseudorapidity of
    the decay muon, $\eta^{\mu}$. Because of the $V\!-\!A$ coupling, a $W$ boson
    produced in the direction of the incoming quark is left-handed in the helicity
    frame, so this fraction is a proxy for the left-handed fraction. The
    distributions are obtained from the generator-level
    \textsc{MadGraph5\_aMC@NLO} sample at $\sqrt{s}=13\,\mathrm{TeV}$, without
    detector simulation or selection requirements; events with no quark or
    antiquark in the initial state are excluded. As a function of $y_{W}$, the
    fraction rises from 0.5 at central rapidity, where the two beams are
    symmetric, to above 0.95 for $y_{W}\gtrsim4$, almost identically for both
    charges. The shaded band indicates the approximate LHCb acceptance. As a
    function of $\eta^{\mu}$, the two charges diverge. Left-handed $W^{-}$ bosons
    emit the $\mu^{-}$ preferentially along the $W$ flight direction, so the
    fraction for $W^{-}$ increases towards 1 at large $\eta^{\mu}$. Left-handed
    $W^{+}$ bosons emit the $\mu^{+}$ preferentially against it, so the fraction
    for $W^{+}$ is approximately 0.7 for $\eta^{\mu}\lesssim3.5$ and falls to 0.4 at
    $\eta^{\mu}=5$. The helicity composition seen in the measured muon kinematics
    therefore differs from that in the $W$-boson kinematics. The statistical
    uncertainties are smaller than the marker size.}
\label{fig:alignment}
    \end{center}
\end{figure}
\clearpage

\addcontentsline{toc}{section}{References}
\bibliographystyle{LHCb/LHCb}
\bibliography{main,LHCb/standard,LHCb/LHCb-PAPER,LHCb/LHCb-CONF,LHCb/LHCb-DP,LHCb/LHCb-TDR}

\newpage
\centerline
{\large\bf LHCb collaboration}
\begin
{flushleft}
\small
R.~Aaij$^{40}$\lhcborcid{0000-0003-0533-1952},
M.~Abdelfatah$^{72}$,
A.S.W.~Abdelmotteleb$^{60}$\lhcborcid{0000-0001-7905-0542},
C.~Abellan~Beteta$^{54}$\lhcborcid{0009-0009-0869-6798},
F.~Abudin\'en$^{62}$\lhcborcid{0000-0002-6737-3528},
T.~Ackernley$^{64}$\lhcborcid{0000-0002-5951-3498},
A.A.~Adefisoye$^{72}$\lhcborcid{0000-0003-2448-1550},
B.~Adeva$^{50}$\lhcborcid{0000-0001-9756-3712},
M.~Adinolfi$^{58}$\lhcborcid{0000-0002-1326-1264},
P.~Adlarson$^{88,45}$\lhcborcid{0000-0001-6280-3851}
% ...

{\footnotesize \it

% ...
$^{40}$Nikhef National Institute for Subatomic Physics, Amsterdam, Netherlands\\
% ...
$^{45}$National Center for Nuclear Research (NCBJ), Warsaw, Poland\\
% ...
$^{50}$Instituto Galego de F{\'\i}sica de Altas Enerx{\'\i}as (IGFAE), Universidade de Santiago de Compostela, Santiago de Compostela, Spain\\
% ...
$^{54}$Physik-Institut, Universit{\"a}t Z{\"u}rich, Z{\"u}rich, Switzerland\\
% ...
$^{58}$H.H. Wills Physics Laboratory, University of Bristol, Bristol, United Kingdom\\
% ...
$^{60}$Department of Physics, University of Warwick, Coventry, United Kingdom\\
% ...
$^{62}$School of Physics and Astronomy, University of Edinburgh, Edinburgh, United Kingdom\\
$^{63}$School of Physics and Astronomy, University of Glasgow, Glasgow, United Kingdom\\
$^{64}$Oliver Lodge Laboratory, University of Liverpool, Liverpool, United Kingdom\\
% ...
$^{72}$Syracuse University, Syracuse, NY, United States\\
% ...
$^{88}$Department of Physics and Astronomy, Uppsala University, Uppsala, Sweden, associated to $^{63}$\\
% ...
}
\end{flushleft}

\end{document}

\or

  \input{LHCbInternal/main-prl.tex}

\or

  \def\PRLwordcount{1}
  \input{LHCbInternal/main-prl.tex}

\else

  \errmessage{Unknown build mode}

\fi